\documentclass[12pt]{article}
\pdfoutput=1
\usepackage[utf8]{inputenc}
\usepackage[english]{babel}
\usepackage{amsmath,bbm}

\numberwithin{equation}{section}

\usepackage{hyperref}
\usepackage{amsfonts}
\usepackage{amssymb}
\usepackage{graphicx}
\newtheorem{theorem}{Theorem}
\newtheorem{lemma}{Lemma}

\newtheorem{proposition}{Proposition}
\newtheorem{definition}{Definition}

\newcommand{\beq}{\begin{equation}}
\newcommand{\eeq}{\end{equation}}
\newcommand{\bea}{\begin{eqnarray}}
\newcommand{\eea}{\end{eqnarray}}
\newcommand{\prf}{{\noindent \bf Proof\; \; }}
\newcommand{\qed}{{\hfill $\Box$}}

\newcommand{\tr}{{\rm tr\,}}

\newcommand{\Tr}{{\rm\bf Tr\,}}

\newcommand{\cP}{{\cal P}}
\newcommand{\cJ}{{\cal J}}
\newcommand{\cT}{{\cal T}}
\newcommand{\cV}{{\cal V}}
\newcommand{\cL}{{\cal L}}
\newcommand{\cE}{{\cal E}}
\newcommand{\cF}{{\cal F}}
\newcommand{\cG}{{\mathcal G}}
\newcommand{\cH}{{\mathcal H}}
\newcommand{\cA}{{\mathcal A}}

\usepackage[left=2cm,right=2cm,top=2cm,bottom=2cm]{geometry}

\begin{document}

\begin{center}
\textbf{{\large Renormalization of an Abelian Tensor Group Field Theory:\\
\medskip
Solution at Leading Order}}
\end{center}

\begin{center}
\vspace{20pt}
Vincent Lahoche\footnote{vincent.lahoche@th.u-psud.fr; LPT-UMR 8627, Universit\'e Paris 11, 91405 Orsay Cedex, France, EU.}, 
Daniele Oriti\footnote{daniele.oriti@aei.mpg.de, Max Planck Institute for Gravitational Physics, Albert Einstein Institute,
 Am M\"uhlenberg 1, 14476, Potsdam, Germany}, Vincent Rivasseau
\footnote{vincent.rivasseau@th.u-psud.fr; Laboratoire de physique th\'eorique, UMR 8627, Universit\'e Paris 11, 91405 Orsay Cedex, France, and
Perimeter Institute for Theoretical Physics, 31 Caroline St. N, N2L 2Y5, Waterloo, ON, Canada}


\begin{abstract}
We study a just-renormalizable tensorial group field theory of rank six with quartic melonic interactions and Abelian group $U(1)$. We introduce the formalism of the intermediate field, which allows a precise characterization of the leading order Feynman graphs. We define the renormalization of the model, compute its (perturbative) renormalization group flow and write its expansion in terms of effective couplings. We then establish closed equations for the 
two point and four point functions at leading (melonic) order. Using the effective expansion and its uniform exponential bounds 
we prove that these equations admit a unique solution at small renormalized coupling. 

\medskip
\noindent Pacs numbers: 11.10.Gh, 04.60.-m, 02.10.Ox
\noindent Report numbers: AEI-2014-058; ICMPA/MPA/2014/23

\end{abstract}
\end{center}



\section{Introduction}

Tensor group field theory (hereafter TGFT)  is a background-independent formalism for quantum gravity. Using  the powerful quantum field theory language, it offers both a tentative definition of the fundamental degrees of freedom of quantum spacetime and a precise encoding of their quantum dynamics. It combines the results of tensor models \cite{review,Rivasseau:2011hm} about the combinatorics of random discrete spaces and the insights of loop quantum gravity \cite{LQG} about quantum geometry. More in detail, TGFTs are quantum field theories on Lie groups, characterized by a peculiar non-local pairing of field arguments in their interactions, whose immediate consequence is that their Feynman diagrams are dual to cellular complexes rather than simple graphs. The quantum dynamics is thus defined, in perturbation theory, by a sum over such cellular complexes (interpreted as discrete spacetimes) weighted by model-dependent amplitudes, in turn functions of group-theoretic data. 
Historically, group field theories (GFTs) \cite{boulatov,GFTreviews} grew out of tensor models for 3d and 4d gravity \cite{tensor}, themselves a generalization of the matrix model definition of 2d Riemannian quantum gravity \cite{matrix}. In tensor models, the dynamics of a quantum spacetime is given by a sum over equilateral d-dimensional triangulations, generated as the Feynman expansion of the partition function for a finite rank-d tensor, and weighted by (the equilateral restriction of) the Regge action for simplicial gravity. They are thus prototypical models of purely combinatorial random geometries. GFTs arise when the domain of the tensors is extended to a group manifold, and the first models \cite{boulatov} make use of these additional data to define amplitudes corresponding to state sum models of topological BF theory (by incorporating appropriate gauge invariance conditions, to which we will return in the following). Soon it was realized \cite{carloPR} that these group-theoretic data gave the boundary states of the same models the structure of loop quantum gravity states \cite{LQG}. Later \cite{DP-F-K-R}, indeed, GFTs were shown to provide a complete definition of the dynamics of the same quantum states as their Feynman amplitudes are given by spin foam models \cite{SF}, a covariant definition of the dynamics of LQG spin networks, in turn dual to simplicial gravity path integrals \cite{danielearistide}. Now, they are understood as a natural second quantized formulation of loop quantum gravity \cite{GFT-LQG}, and GFT models incorporating more quantum geometric features of LQG states and simplicial geometry are indeed among the most interesting ones.

In the meantime, tensor models have witnessed an important resurgence, in the form of {\it colored tensor models} \cite{review, color}. These solved many issues raised by earlier tensor models and allowed a wealth of important mathematical results to be obtained. They triangulate pseudo-manifolds with only local singularities \cite{lost}, having in particular no \emph{tadfaces} (i.e. a face which runs several times through a single edge). Most importantly, they admit a large $N$ expansion \cite{expansion} (where $N^d$ is the size of the tensor), whose leading order is now well understood. The leading graphs in this limit, the \emph{melonic} graphs, form particularly simple ``stacked" triangulations of the sphere in any dimension \cite{critical}. Their appearance is a very general phenomenon \cite{universality,uncoloring}. Some of these results have immediately been extended to topological GFTs and multiorientable models \cite{Tanasa:2011ur}, and beyond the leading order, to define interesting double scaling limits \cite{doublescal}. 
	
Incorporating the insights of colored tensor models into GFTs leads to TGFTs. Here, the GFT fields are required to transform as proper tensors under unitary transformations and their interactions are required to have the additional $ U(N)^{\otimes d}$ invariance, which can be interpreted as a new notion of locality, hence singles out a new \emph{theory space} \cite{vincentTS}. In turn, this invariance requires their arguments to be labeled (ordered).  Both facts are crucial for GFT renormalization. 

GFT renormalization is in fact a thriving area of current research. Given that the first definition of the GFT quantum dynamics is in terms of a perturbative expansion around the Fock vacuum, the first aim is to prove renormalizability of specific models, showing therefore their consistency as quantum theories. Second, one is interested in unraveling the phase space of GFT models, looking in particular for a phase in which approximate smooth geometric physics (governed by some possibly modified version of General Relativity) emerges from the collective behavior of their pre-geometric degrees of freedom \cite{danieleemergence}, maybe through a process of condensation. The search for such a geometric phase, and the associated phase transition(s), is common to tensor models \cite{critical}, loop quantum gravity \cite{Tim-DGP}, and spin foam models \cite{biancabenny}, but also to other related approaches like (causal) dynamical triangulations \cite{CDTphasetrans}. Moreover, it has been conjectured \cite{danieleemergence} to have a direct physical interpretation in a cosmological context \cite{LeeJoaoContaldi}, and some recent results in GFT support this conjecture  \cite{GFTcondensate}. 
	
The TGFT framework is well-suited for renormalization, as one can import more or less standard QFT techniques even in such background independent context. One ingredient is the new notion of locality provided by the $U(N)^{\otimes d}$ invariance of tensor interactions. The other ingredient, a notion of {\it scale} is naturally assumed to be given by the decomposition of GFT fields in group representation. This is fully justified in terms of spectra of the kinetic operator (as in standard QFT) when a Laplacian on the group manifold is used, as suggested by the analysis of radiative corrections to topological GFT models \cite{Geloun:2011cy} (which correspond to ultra-local truncations of truly propagating models). All these ingredients, it turns out, speak to one another very nicely, as indeed in TGFT models counter-terms necessary to cure divergences remain of the same form of the initial interactions. More precisely, by precise power counting of divergences, one sees that at large ultraviolet (UV) scales (in the sense of large eigenvalues of the group Laplacian) connected subgraphs which require renormalization seem local (as defined by tensor invariance) when observed at lower scales. 

A large amount of results has been already obtained. For models without gauge invariance the proof of renormalizability at all orders, which started with \cite{BGR}, now includes a preliminary classification
of renormalizable models \cite{GFTRenormClass} and studies of the equations they satisfy \cite{GFTRenormFur}. Then Abelian \cite{Carrozza12013,Sam1} and non-Abelian gauge invariance  
(whose important role we already emphasized) has been included \cite{Carrozza22013,Carrozza2014}. The computations of beta functions typically shows UV asymptotic freedom 
\cite{BG1,Samary:2013xla} to be 
a rather generic feature of TGFTs, even if the analysis of more involved models is in fact quite subtle \cite{Sylvain}. Renormalizability and UV asymptotic freedom are the two key properties of non-Abelian gauge theories which form the backbone of the quantization of all physical interactions except gravity, hence it is encouraging to find them also in TGFTs, which aim at quantizing gravity. 

Once renormalizability (and possibly asymptotic freedom) is established, the next stage is to understand the \emph{infrared} (IR) behavior of the renormalization group flow,
in particular phase diagrams and phase transitions.
One can prove that the leading ``melonic" order of tensor models and of topological GFTs  exhibits a phase transition, 
corresponding to a singularity of the free energy for a certain value of the coupling \cite{critical,Baratin:2013rja}. 
The critical susceptibility can be computed at least for simple tensor models to be equal to $1/2$. In the same tensor models context, in which the only notion of distance is the graph distance, one sees a phase corresponding to branched polymers, with Hausdorff dimension 2 and spectral dimension $4/3$ 
\cite{branched}, as in CDT. In GFTs and TGFTs, where the group theoretic data play a prominent role, not only computing observables and critical exponents, but also finding the nature of the transitions and their physical interpretation is much more difficult. 

Therefore we need more analytic tools. One powerful scheme is provided by functional renormalization techniques. These have been developed for TGFTs for the first time in \cite{Benedetti:2014qsa}. Applied to the (comparatively) easy case of an Abelian rank-3 model, the RG flow equations could be derived and the phase diagram be plotted in the key UV and IR regimes, 
showing evidence for a phase transition to a condensed phase, at least in some approximation. 
%
%

In this paper we perform a leading order analysis of the correlation functions of a simple TGFT
with quartic melonic interactions and U(1) group, in dimension 6, endowed with gauge invariance conditions. This model is just-renormalizable \cite{Sam1}, and asymptotically free \cite{Samary:2013xla}.
Hence it should exist at the level of constructive field theory \cite{Rivasseau:1991ub} (see \cite{Delepouve:2014hfa} for the construction
of a simpler super-renormalizable TGFT). Although we shall not achieve such a complete non-perturbative analysis in this paper,
we provide some significant steps in this direction. We define the intermediate field formalism for our model and with a multi-scale analysis we establish its renormalizability, compute the beta function of the model and check its asymptotic freedom. In this way we recover all the results of \cite{Sam1} and \cite{Samary:2013xla}. The development of the intermediate field method for our model is in itself, we believe, an interesting result. It is known to be particularly convenient for quartic tensor models \cite{Gurau:2013pca,Delepouve:2014bma}, and should become a standard tool for TGFT's as well. One should notice in particular that in our case, due to the gauge conditions, the intermediate fields are of a 
vector rather than matrix type, a promising new feature.

We then define the effective expansion of the model, which sits ``in between" the bare and the renormalized expansion. Its 
main advantage is to be free of \emph{renormalons} \cite{Rivasseau:1991ub}.  We check this fact again in our model by establishing uniform exponential upper bounds on effective amplitudes.
We also establish closed equations for the leading order (i.e. melonic approximation) to the two-point and four-point functions. Combining all these results
proves that these closed equations admit a unique solution for small enough renormalized coupling, and
gives full control over the melonic approximation of the theory, bringing it to the level of analysis of the Grosse-Wulkenhaar 
non-commutative field theory \cite{Grosse:2004yu}.

Similar closed equations have been written for another renormalizable TGFT theory, in dimension 5 and with a simpler propagator 
without gauge invariance conditions in \cite{ousmane}. The renormalization and numerical analysis of these equations
have been recently developed in \cite{Samary:2014oya}.


Our paper is organized as follows. In Section 2 we define the model and its intermediate field representation.
In Section 3 we establish and analyse its power-counting with multi-scale analysis. Section 4 describes its renormalization, computes the beta function (in agreement with 
\cite{Samary:2013xla}), introduces the effective expansion and establishes uniform bounds on the corresponding effective amplitudes. 
Section 5 writes the closed equations for the melonic approximation to the bare and renormalized two point and four-point functions, and
completes the proof that these equations have a unique solution at small renormalized coupling, which is 
in fact the Borel sum of their renormalized expansion.

\section{The Model}

In this section, we shall briefly recall the basics of TGFTs models with closure constraint (gauge invariance) and Laplacian propagator. Then we shall focus on a particular $U(1)$ quartic model at rank six first defined in \cite{Carrozza12013}. Within this section definitions and computations are still formal since
we do not introduce cutoffs; this will be done in the next sections.

\subsection{General Formalism for TGFTs}
A generic TGFT is a statistical field theory for a tensorial field, for which the entries are living in a Lie group $G$, generally compact, such as $U(1)$ or $SU2)$ for the simplest cases. A family of such models was defined and renormalized to all orders in \cite{Carrozza12013,Carrozza22013,Sam1,Carrozza2014}\footnote{Renormalizability has not been yet established for models based on the Lorentz group, which is non-compact. However, at least intuitively, one could expect the additional difficulties present in the non-compact case to be rather of IR nature than of UV nature, from the point of view of TGFT renormalizability; this would imply similar renormalizability results as in the compact group case.}. 
	
The theory is defined by an action and by the following partition function
\begin{equation}
S(\bar{\phi},\phi)=S_{int}(\bar{\phi},\phi)- \bar{J} \cdot \phi - \bar{\phi} \cdot J, \qquad
Z(\bar J, J)=\int{d\mu_{C}(\bar{\phi},\phi) e^{-S[\bar{\phi},\phi]}},
\end{equation}
where $S_{int}$ is the interaction and $d\mu_{C}$ is a Gaussian measure characterized by its covariance $C$. The fields $\phi$ and $\bar{\phi}$ are complex functions $\bar{\phi}, \phi: G^{d} \mapsto  \mathbb{C}$
noted $\phi(g_1, \cdots,  g_d) = \phi(\vec g)$,
$ \vec{g}=(g_{1},g_{2},...,g_{d})$, and  $\bar\phi(\vec g\,')$, $ \vec{g}\,'=(g'_{1},g'_{2},...,g'_{d})$. 
They should equivalently be also considered as rank-$d$ tensors, that is elements of the tensor space $L^2 (G)^{\otimes d}$, where $L^2 (G)$ is the space
of functions on $G$ which are square-integrable  with respect to the Haar measure. 
The $2N$-point Green functions are obtained by deriving $N$ times  with respect to sources $J$ and $N$ times  with respect to anti-sources $\bar J$
\bea
G_{2N}(\vec g_1,\cdots, \vec g_{N}, \vec g\,'_1 ,  \cdots \vec g\,'_N)
=\frac{\partial^{2N} Z(\bar J, J)}{\partial J_1 (\vec g_1) \partial \bar J_1 (\vec g\,'_1)
\cdots \partial J_N (\vec g_N)\partial \bar J_N (\vec g\,'_N) }\Big|_{{ J}={\bar J}=0}.
\eea

The Gaussian measure is defined by the choice of the action's kinetic term. TGFTs 
such as those of \cite{BGR,GFTRenormClass} use a mass term plus the canonical Laplace-Beltrami operator $\Delta$
on the group $G^d$, hence correspond to the formal normalized measure
\begin{equation}
d\mu_{C_{0}}(\bar{\phi},\phi) = \frac{1}{Z_0}  e^{-S_{kin}[\bar{\phi},\phi]}  D \bar\phi D \phi \label{Leb}
\end{equation}
with
\begin{equation} \label{lapbel}
S_{kin}(\bar{\phi},\phi)=\int{[dg]^{d} \bar{\phi}(\vec{g})[(-\Delta +m^{2})\phi](\vec{g})} , \quad
\end{equation}
where $dg$ is the Haar measure on the group. Although the Lebesgue measure $D \bar\phi D \phi$  in \eqref{Leb} is ill-defined,
the measure $d\mu_{C_0}$ itself is well-defined, and the propagator $C_0$ in the parametric (or Schwinger) representation is
\begin{equation}
C_{0}(\vec{g},\vec{g\,''})=\int_{}^{} \mathrm{d}\mu_{C_{0}}(\bar{\phi},\phi) \bar{\phi}(\vec{g})\phi(\vec{g}\,')=\int_{0}^{\infty} \mathrm{d}\alpha e^{-\alpha m^{2}} \prod_{c=1}^{d} K_{\alpha}(g_{c}g{'}_{c}^{-1}),
\end{equation}
where $K_{\alpha}$ is the heat kernel associated to the Laplacian operator, and $c$ is our generic notation for a color index running from 1 to $d$. 
In momentum space this propagator becomes diagonal.  Let us from now on restrict to the case $G= U(1)$. The Fourier dual of $U(1)$ is ${\mathbb Z}$, hence in momentum space, we note $\vec p = (p_1, \cdots  , p_d)\in {\mathbb Z}^d$, where $p_c \in {\mathbb Z}$ is called the strand momentum
of color $c$, and
we have
\begin{equation}
C_{0}(\vec{p},\vec{p}\,')=\prod_{c=1}^d \delta (p_c, p'_c) \dfrac{1}{\vec{p}\,^2+m^2}.
\end{equation}

In the specific TGFT we study in this paper, we want the field configurations to obey the additional gauge invariance 
\begin{equation}
\phi(g_{1},g_{2},...,g_{d})=\phi(hg_{1},hg_{2},...,hg_{d}), \; \bar \phi(g_{1},g_{2},...,g_{d})=\bar \phi(hg_{1},hg_{2},...,hg_{d}) \quad  \forall h\in G .
\end{equation}
This gauge invariance complicates slightly the writing of the model. In order to implement it, we could introduce the (idempotent) projector $P$ which projects the fields on the subspace of gauge-invariant fields, 
then equip the interaction vertices and propagators with such projectors. But in this case the tensorial symmetry
$U(N)^{\otimes D}$ symmetry of the interaction vertex (which provides the analog of a locality principle for renormalization)
would be blurred. Hence the best solution, used in \cite{Carrozza12013}, consists in implementing the gauge invariance 
directly on the Gaussian measure by introducing a group-averaged covariance
\begin{equation}
C(\vec{g},\vec{g}\,')=\int_{}^{} \mathrm{d}\mu_{C}(\bar{\phi},\phi) \bar{\phi}(\vec{g})\phi(\vec{g}\,')=
\int_{0}^{\infty} \mathrm{d}\alpha e^{-\alpha m^{2}} \int{dh \prod_{c=1}^{d} K_{\alpha}(g_{c}hg{'}_{c}^{-1})} .
\end{equation}
In other words, we introduce the gauge invariance projector $P$ only in the propagator of the theory\footnote{Additional insertions of $P$ on the  vertex would result in the same Feynman amplitudes, since $P^2 = P$.}. In momentum space we have.  
\begin{equation}\label{u0covariance}
C(\vec{p},\vec{p}\,')=  \prod_{c=1}^d   \delta (p_c, p'_c) \dfrac{\delta \left( \sum_{c}p_c\right)}{\vec{p}^2+m^2}.  
\end{equation}
From now on we shall remember that the covariance is diagonal in momentum space, with diagonal values 
\begin{equation}\label{u1covariance}
C(\vec{p})=  \dfrac{\delta \left( \sum_{c}p_c\right)}{\vec{p}^2+m^2}. 
\end{equation}
hence defining the set $\cP = \{ \vec p \in {\mathbb Z}^6  \: \vert \;   \sum_{c} p_c  =0 \} $ of momenta
satisfying the gauge constraint, all Green functions of our theory can in fact be defined for restricted momenta $\vec p \in \cP$,
or if one prefers, are zero outside $\cP$.

\begin{figure}[h]
\begin{center}
\includegraphics[scale=0.7]{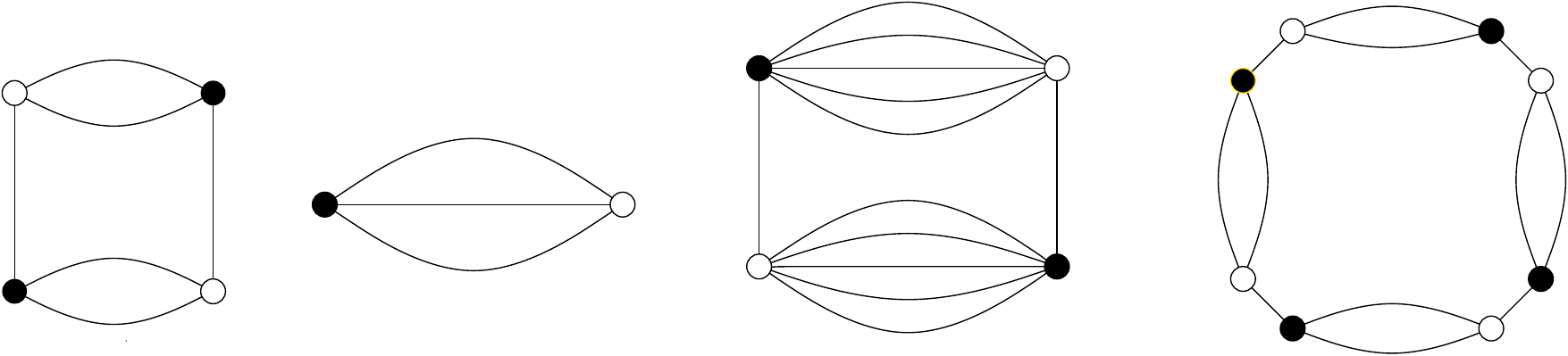} 
\caption{Some connected tensor invariants}\label{figinv}
\end{center}
\end{figure}

TGFT interactions by definition belong to the tensor theory space \cite{universality,uncoloring,vincentTS} spanned by $U(N)^{\otimes d}$ invariants. Hence 
the most general polynomial interaction is a sum over a finite set $\mathcal{B}$ of such invariants $b$, also called $d$-bubbles, associated with different coupling constants $t_b$
\begin{equation}
S_{int}(\bar{\phi},\phi)=\sum_{b\in \mathcal{B}}t_{b}I_{b}(\bar{\phi},\phi),
\end{equation}
where  $I_{b}$ is the connected invariant labeled by the bubble $b$. Graphically, each bubble is associated with a bipartite 
$d$-regular edge-colored graph. Each color $c \in \{1,2,..,d\}$ is associated with a half-line at each vertex, and each vertex
bears respectively a field $\phi$ or its complex conjugate $\bar \phi$ according to its black or white color. 
The edge coloring of the bipartite graph allows to visualize the $U(N)^{\otimes d}$ invariance by showing the exact pairing of fields and anti-fields
argument of the same color. 
Such graphs also enable to visualize whether the interaction is \emph{connected} or not. Some examples of connected invariants at ranks $d= 3$  and 
$d=6$ are shown in Figure \ref{figinv}.

\begin{figure}[here]
\begin{center}
\includegraphics[scale=0.45]{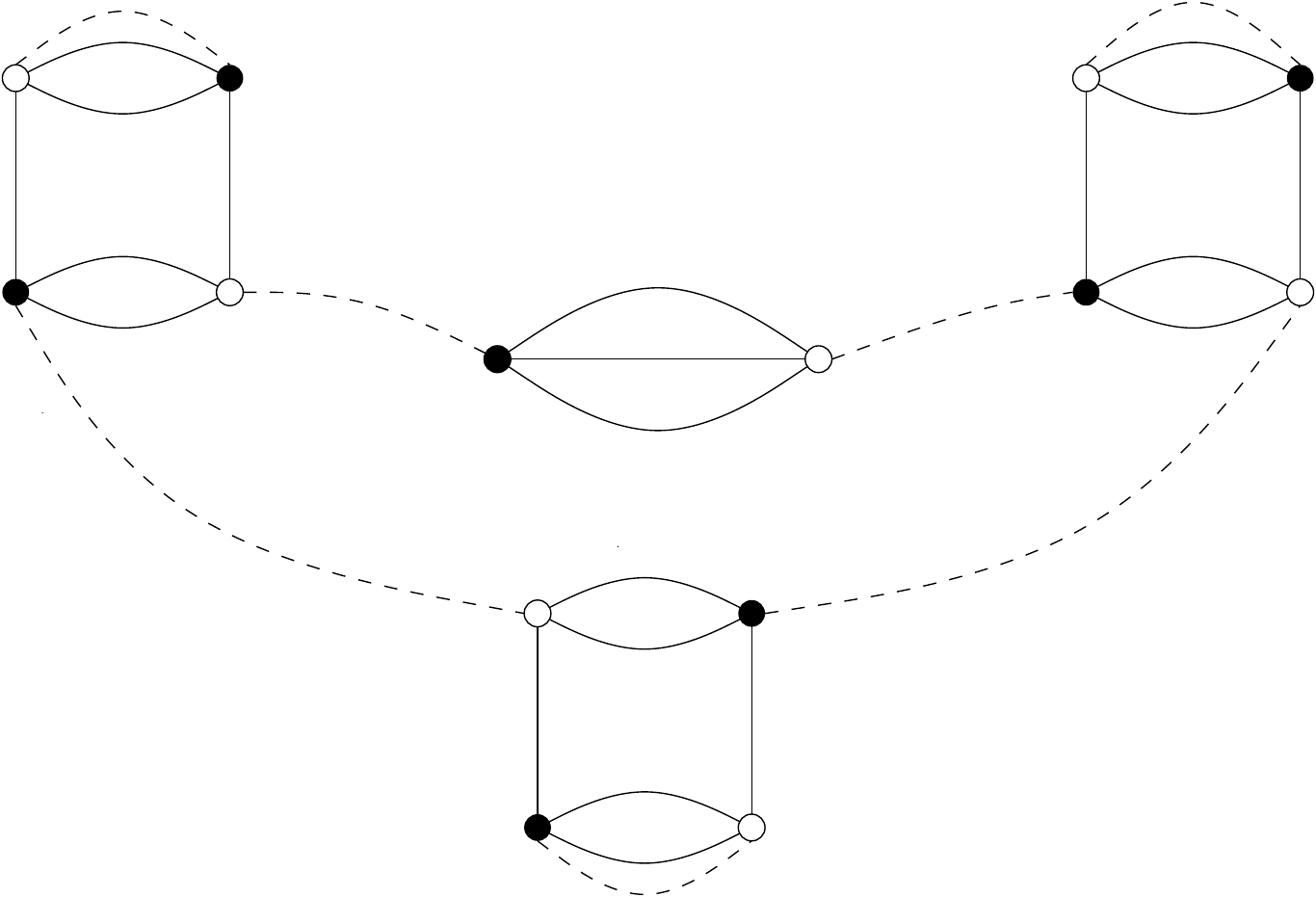}
\caption{A tensorial vacuum (N=0) rank-three Feynman graph}\label{figtens}
\end{center}
\end{figure}

The Feynman amplitudes of the perturbative expansion are associated with Feynman graphs whose vertices belong to the set $\mathcal{B}$ of the interaction 
$d$-bubbles. A Wick contraction is represented
by a dotted line. Figure \ref{figtens} gives an explicit example for $d=3$. 

For a Feynman graph  $\mathcal{G}$, we note $\cV(\mathcal{G})$,  $\cL(\mathcal{G})$ and $\cE(\mathcal{G})$
the sets of the vertices (the $d$-bubbles), internal (dotted) lines and external (dotted) half-lines, and $V(\mathcal{G})$,  
$L(\mathcal{G})$ and $E(\mathcal{G})=2N (\mathcal{G})$ the number of elements in these sets. The number of vertices $V$ is also identified
with the order of perturbation, also often noted $n$.

The Green functions are given by a sum over Feynman graphs (connected or not)
\begin{equation}
G_{2N}=\sum_{\mathcal{G}  , \; E(\mathcal{G})=2N}\;\;  \dfrac{1}{s(\mathcal{G})}\left( \prod_{b\in \mathcal{B}}(-t_{b})^{n_{b}(\mathcal{G})} \right)  \mathcal{A}_{\mathcal{G}},
\label{greenfey}
\end{equation}
where $n_b$ is the number of vertices of type $b$ and $s(\mathcal{G})$ is the graph symmetry factor (dimension of the automorphism group). Note that 
expanding each vertex $b$ as a $d$-regular bipartite edge-colored graph as in Figure \ref{figinv} and coloring the dotted lines
with a new color 0, any such graph $\mathcal{G}$ 
is therefore canonically associated to a unique $(d+1)$-regular bipartite edge-colored graph, for which the vertices are the black and white nodes, as shown in 
Figure \ref{figtens}. 
Hence it defines an associated $d$-complex, in which in particular \emph{faces} are easily defined as the bi-colored connected components \cite{universality,uncoloring}.
These faces are either closed or \emph{open} if they end up on external half-lines. 

The connected Green functions or cumulants $G^c_{2N}$ are obtained by restricting sums such as \eqref{greenfey} to connected graphs $\mathcal{G}$, and are obtained from the generating functional 
\bea
W(\bar J, J) = \log [Z(\bar J, J) ]
\eea through 
\bea
G^c_{2N}(\vec g_1,\cdots, \vec g_{N}, \vec g\,'_1 ,  \cdots \vec g\,'_N)
=\frac{\partial^{2N} W(\bar J, J)}{\partial J_1 (\vec g_1) \partial \bar J_1 (\vec g\,'_1)
\cdots \partial J_N (\vec g_N)\partial \bar J_N (\vec g\,'_N) }\Big|_{{ J}={\bar J}=0}.
\eea
The vertex functions $\Gamma_{2N}$ are obtained by restricting sums such as \eqref{greenfey} to one particle irreducible
amputated graphs $\mathcal{G}$ (amputation mean we replace all the external propagators for dotted half-lines by 1). They are the coefficients
of the Legendre transform of $W (\bar J, J)$.

Using the convolution properties of the heat kernel (following from the composition properties of its random path representation), the Feynman  amplitude $\mathcal{A}_{\mathcal{G}}$ of $\mathcal{G}$ can be expressed in direct space as \cite{Carrozza2014}
\begin{align}
\mathcal{A}_{\mathcal{G}}&= {\left[ \prod_{\ell \in \mathcal{L}(\mathcal{G})} \int_{0}^{\infty}{d\alpha_{\ell} e^{-\alpha_{\ell} m^{2}}} \int {dh_{\ell}} \right]}{\left( \prod_{f\in \cF(\mathcal{G})} K_{\alpha_{(f)}} {\left( \vec{\prod}_{\ell\in\partial{f}}h_{\ell}^{\epsilon_{\ell f}} \right)} \right)} \times \nonumber \\
&\quad{\left( \prod_{f\in \cF_{ext}(\mathcal{G})} K_{\alpha_{(f)}} {\left( g_{s(f)}\vec{\prod}_{\ell\in\partial{f}}h_{\ell}^{\epsilon_{\ell f}}g^{-1}_{t(f)} \right)} \right)}.
\end{align}
In this expression, $\cF({\mathcal{G}})$ is the set of internal faces of the graph, $\cF_{ext}(\mathcal{G})$ the set of external faces, and $\epsilon_{\ell f}$ the adjacency matrix which is non zero if and only if  the line $\ell$ belongs to the face $f$ and is $\pm 1$ according to their relative orientation. 
We noted $\alpha(f)=\sum_{\ell \in\partial f} \alpha_{\ell}$ the sum of Schwinger parameters along the boundaries-lines of the face $f$, and $g_{s(f)}$ or $g_{t(f)}$ the boundary variables in the open face $f$, $s$ for ``source" and $t$ for ``target" variables. We use also the notation $\cF$ for the set of faces and $F$ for its cardinal (number of elements).

These amplitudes $\mathcal{A}_{\mathcal{G}}$ can be interpreted as lattice gauge theories defined on the cellular complexes dual to the Feynman diagrams $\mathcal{G}$. The group elements $h_\ell$ (resp. $g_{s(f)}$, $g_{t(f)}$) define a discrete gauge connection associated to the edges $\ell$ (resp. boundary edges) of the cellular complex, and the ordered products $\vec{\prod}_{\ell\in\partial{f}}h_{\ell}^{\epsilon_{\ell f}}$ (resp. $g_{s(f)}\vec{\prod}_{e\in\partial{f}}h_{e}^{\epsilon_{ef}}g^{-1}_{t(f)}$) are its holonomies (discrete curvature) associated to bulk (resp. boundary) faces of the same complex\footnote{
In models of 4d quantum gravity that bear a closer relation with loop quantum gravity, and that encode more extensively features of simplicial geometry, additional conditions called \emph{simplicity constraints} are imposed \cite{SF,danielearistide,GFT-LQG}. Obviously, they complicate the structure of the amplitudes, making them richer. We do not consider these additional constraints here.}.

Due  to the diagonal character of the propagator in momentum space, these Feynman amplitudes are easier to express in the momentum representation.
In particular the momentum conservation along faces due to the $\delta $ functions in \eqref{u0covariance} ensures that when expressed in momentum space
non-zero Green functions of the theory of order $2N$ must themselves develop into sums over $U(N)^{\otimes d}$
tensor-invariants of the momenta of order $N$; 
in other words to any entering momentum $p_c$ must correspond an exiting momentum with same value $p'_c = p_c$.
In particular the two point function in momentum space is a function $G_2 (\vec p)$ of a single momentum $\vec p \in {\mathbb Z}^d$, and the 
connected four point function $G_4^c$ is a sum over all quartic invariants of the theory.
In general the contribution of a given specific tensor invariant is complicated to extract from the Green functions. It
requires a somewhat subtle decomposition using Weingarten functions,
which we shall not detail here, referring the reader to  \cite{Gurau:2013pca,Delepouve:2014bma}.

\subsection{The Quartic Melonic U(1)-model in dimension 6}

After this quick overview of general TGFTs, we come to the particular model studied in this paper, namely the $d=6$ Abelian quartic model with melonic interactions. 
It is the simplest just-renormalizable model (with no simplicity constraints) in the classification of gauge invariant TGFT models \cite{Carrozza22013}. As such, it is  also the simplest interesting testing ground for the analytic techniques we develop here.  

\begin{figure}[here]
\begin{center}
\includegraphics[scale=0.7]{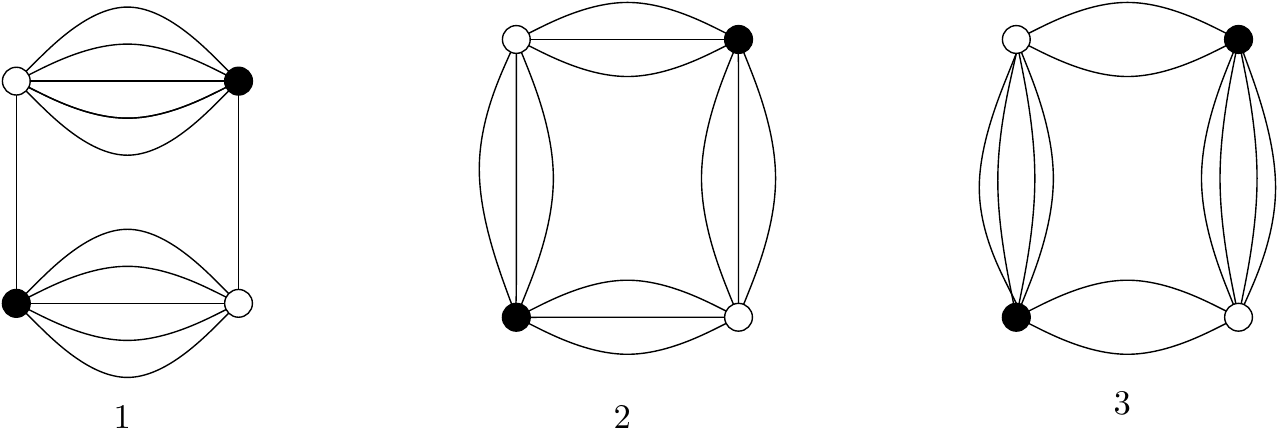} 
\caption{The quartic tensor interactions at rank 6}\label{figquartic6}
\end{center}
\end{figure}

General quartic interactions at rank 6 are of the three types indicated in Figure \ref{figquartic6}. Melonic interactions correspond to the type 1. They are leading in the $1/N$ tensorial expansion and are marginal in the renormalization group (RG) sense, the other ones being irrelevant. 

Hence the interaction part of the action considered from now on is the sum of all the bubbles of type 1. 
There are 6 of them, characterized by a unique index $c$ referring to the special color which colors the two lonely lines of the bubble:
\begin{equation}
S_{int}=\sum_{c=1}^{6}{\lambda_{c} {\rm Tr}_{b_{c}}(\bar{\phi}\phi)} . \label{ccc}
\end{equation}
More explicitly a quartic interaction $b_1$ with special color 1 writes
\begin{eqnarray}
{\rm Tr}_{b_{1}}(\bar{\phi}\phi)&=&\int{d\vec{g} d\vec{g}\,'} \, \bar{\phi} (g_{1}, g_2, \cdots, g_6)  \phi (g'_{1}, g_2, \cdots, g_6) 
\bar{\phi} (g'_{1}, g'_2, \cdots, g'_6)  \phi (g_{1}, g'_2, \cdots, g'_6) \nonumber \\
&=&\sum_{\vec{p},\vec{p}\,'}   \bar{\phi}(p_{1}, p_2, \cdots, p_6)  \phi (p'_{1}, p_2, \cdots, p_6) 
\bar{\phi}(p'_{1}, p'_2, \cdots, p'_6)  \phi (p_{1}, p'_2, \cdots, p'_6) , \label{traces}
\end{eqnarray}
where the last line is written in Fourier space. Remark that since only fields satisfying the propagator constraints $\sum p_c =0$ can contribute,
in \eqref{traces} we must have $p_1= p'_1$.  Hence each ${\rm Tr}_{b_{c}}$
is a function of fields with 9 (rather than 10) independent strand momenta, because $p_c = p'_c$.
We can therefore in our model simplify \eqref{traces} into 
\begin{equation}
{\rm Tr}_{b_{1}}(\bar{\phi}\phi)=\sum_{\vec p \in \cP, \vec p\,' \in \cP \vert  p_1 = p'_1 }
\bar{\phi}(p_{1}, p_2, \cdots, p_6)  \phi (p_1, p_2, \cdots, p_6)  
\bar{\phi}(p_1, p'_2, \cdots, p'_6)  \phi (p_1, p'_2, \cdots, p'_6). 
\label{traces1}
\end{equation}

From now on we consider only the color-symmetric case $\lambda_c = \lambda$ $\forall c = 1, \cdots, 6$.

As remarked, Green functions in momentum space develop into sums of tensor invariants.
In particular the connected four point  function $G^c_4$ develops over \emph{all} quartic invariants (connected or not).
Hence it develops over the connected invariants of Figure \ref{figquartic6}
and over the disconnected invariant which is the square of the quadratic invariant. This may seem
dangerous at first sight since to be renormalizable our model should not involve in particular renormalization of
invariants of type 2 and 3 which are not part of the initial interaction.

As well known, renormalization is best stated in terms of the vertex functions $\Gamma$. Hence we shall be particularly interested
in computing the two point vertex function or self-energy $\Gamma_2 (\vec p)$ and the four point vertex function $\Gamma_4 (\vec{p}_1,...,\vec{p}_4)$.
These functions are a priori defined on $\cP$ or $\cP^2$. However we shall see that their divergent part is simpler. More precisely we shall
define melonic parts $\Gamma^{melo}_2 (\vec p)$ and $\Gamma^{melo}_4 (\vec{p}_1,...,\vec{p}_4)$ for these vertex functions, and even 
a refined monocolor melonic part $\Gamma^{melo}_{4,mono} (p_c, p_c')$ of $\Gamma^{melo}_4 (\vec{p}_1,...,\vec{p}_4)$, such that
$\Gamma_2 (\vec p)- \Gamma^{melo}_2 (\vec p)$ and $\Gamma_{4,mono} (p_c, p_c')- \Gamma^{melo}_{4,mono} (p_c, p_c')$
are superficially convergent (hence truly convergent after all divergent strict  subgraphs have been renormalized). 
More precisely we shall prove that
\begin{theorem}
There exist two (ultraviolet-divergent) functions $f$ and $g$ of a single strand momentum $p \in {\mathbb Z}$ such that
\begin{equation}
\Gamma^{melo}_2 (\vec p) = -\lambda \sum_{c=1}^6  f(p_c),\;  \Gamma^{melo}_{4,mono} (p_c, p_c') =  -\lambda \delta (p_c , p'_c) g(p_c). \label{goodeq}
\end{equation}
and such that  $\Gamma_2 (\vec p)- \Gamma^{melo}_2 (\vec p)$ and $\Gamma_{4,mono} (p_c, p_c')- \Gamma^{melo}_{4,mono} (p_c, p_c')$
are superficially convergent (hence truly convergent after all divergent strict  subgraphs have been renormalized). All higher order vertex functions are also superficially convergent.
\label{goodth}
\end{theorem}
In particular $\Gamma^{melo}_2$ and $\Gamma^{melo}_{4,mono} (p_c, p_c')$ both
depend in fact of a single non-trivial function, respectively $f$ and $g$, of a single strand momentum in ${\mathbb Z}$. We shall prove 
that the special form \eqref{goodeq} of the primitive divergencies of the theory is compatible with the renormalization of the couplings in \eqref{traces1}. 
In the next section we 
introduce the intermediate field representation in which the functions $f$ and $g$ are particularly simple to 
represent graphically and to compute.

\subsection{The intermediate field formalism}
	
\begin{figure}[here]
\begin{center}
\includegraphics[scale=0.7]{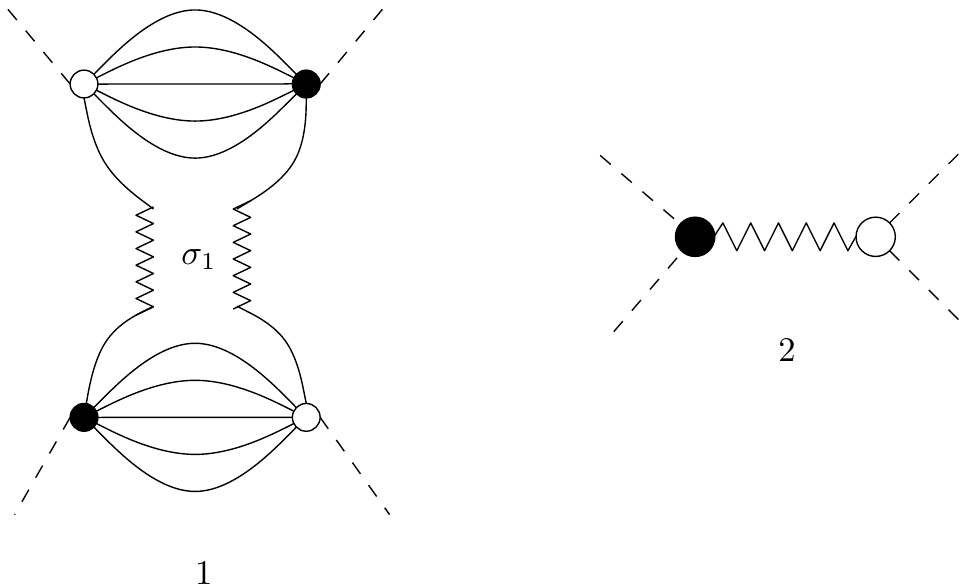} 
\caption{Intermediate field decomposition}\label{figinterdec}
\end{center}
\end{figure}

The intermediate field formalism is a mathematical trick to decompose a quartic interaction in terms of a three-body interaction, 
by introducing an additional field (the intermediate field) in the partition function. It is based on
the well-known property of Gaussian integrals:
\begin{equation}
\int_{-\infty}^{+\infty} dx e^{-x^{2}/2}e^{i\kappa xy}=\sqrt{\pi}e^{-{\kappa}^{2}y^{2}/2}. \label{gaussfourier}
\end{equation}	
We first apply the general method without exploiting gauge invariance, then stress the simplification due to gauge invariance.
This means we start with \eqref{traces} which we want to exhibit as a square. For this we introduce the six auxiliary matrices
$ \sum_{p_2, \cdots, p_6}  \bar\phi (p_1, p_2, \cdots, p_6)  \phi (p'_1, p_2, \cdots, p_6) 
= \mathbb{M}_ {p_{1}, p'_{1} }$, which are quadratic in terms of the initial $\bar \phi$ and $\phi$ and can be thought as 
partial traces over color indices other than 1.
The  interaction in \eqref{traces} can be rewritten as
\begin{equation}
{\rm Tr}_{b_{1}} (\bar{\phi}\phi)= \tr\, {\mathbb{M}^2} \ ,
\end{equation}
where $\tr$ means a simple trace in $\ell^2 ( {\mathbb Z} )$.
Using many times \eqref{gaussfourier} we can decompose this square interaction $\tr \,{\mathbb{M}^2}$ with a new Hermitian matrix $\sigma_{1}$ 
corresponds graphically to ``pinching" the two special strands of color 1 with this matrix field, as indicated in Figure \ref{figinterdec}. More precisely
\begin{equation}
e^{-\lambda \tr({\mathbb{M}^2})}=\frac{\int d{\sigma_{1}}e^{-{\tr(\sigma_{1}^{2})/2}}e^{i\sqrt{2\lambda}\tr({\sigma_{1} \mathbb{M}})}}{\int d{\sigma_{1}}e^{-{\tr(\sigma_{1}^{2})/2}}}.
\end{equation}

The next step is to make this decomposition systematic for the six melonic interactions. Writing
\begin{equation}
\tr({\sigma_{1} \mathbb{M}})= \Tr \bar{\phi}\Sigma^{1} \phi,
\end{equation}
where
\begin{equation}
\Sigma^{1}=\sigma_{1}\otimes \mathbb{I}\otimes\mathbb{I}\otimes\mathbb{I}\otimes\mathbb{I}\otimes\mathbb{I}
\end{equation}
acts in the large tensor space $\ell^2 ( {\mathbb Z} )^{\otimes 6}$ and $\Tr$ means a trace in this large tensor space,
allows to express the previous intermediate field decomposition as
\begin{equation}
e^{\lambda\, \tr\, {\mathbb{M}^2}}=\frac{\int d{\sigma_{1}}e^{-{\tr(\sigma_{1}^{2})/2}}e^{i \sqrt{2\lambda}\Tr \bar{\phi}\Sigma^{1}\phi}}{\int d{\sigma_{1}}e^{-{\tr(\sigma_{1}^{2})/2}}}.
\end{equation}

Using color permutation, we decompose all six bubbles in this way. An intermediate field $\sigma_ {c}$ is therefore associated to 
each quartic bubble $b_c$ with weak color $c$. The operators
\begin{equation} \Sigma^{c} = \mathbb {I} \cdots \otimes \sigma_ {c} \otimes \cdots \mathbb {I} 
\end{equation} commute in the tensor space $\ell^2 ( {\mathbb Z} )^{\otimes 6}$, as they act on different strands.
Introducing $\Sigma = \sum_{c=1}^6 \Sigma^ {c} $, we can rewrite the partition function of the original theory as
\begin{eqnarray}
Z(\bar{J}, J) =\int d\mu_{C}({\phi,\bar{\phi}})  e^{-S_{int}(\bar{\phi},\phi)}&=&\int{d\mu_{C}({\phi,\bar{\phi}}) d \nu ( \sigma) e^{- \bar{\phi} \cdot J - \phi \cdot \bar J  }
e^{i \sqrt{2\lambda}\Tr\bar{\phi}\Sigma^c \phi}}  ,
\end{eqnarray}
the normalized Gaussian measure $d \nu ( \sigma)$ being factorized over colors with trivial covariance identity on each independent coefficient (Gaussian unitary ensemble).
The tensor integral becomes Gaussian, hence can be computed as a determinant. We find:

\begin{proposition}The partition function of the model is given by
\begin{align}
Z(\bar{J}, J) = \int  e^{ - \bar J  (1-i\sqrt{2\lambda}C \Sigma  )^{-1} C J     -\Tr \ln(1-i\sqrt{2\lambda}C \Sigma )  }d \nu ( \sigma)  .  \label{logi}
\end{align}
\end{proposition}
Therefore pairs of sources are become resolvents $(1-{i\sqrt{2\lambda}C \Sigma  })^{-1}$ in this representation. 

Perturbatively one can expand both the interaction logarithm and these resolvents as
\begin{eqnarray}
- \Tr \ln(1-i\sqrt{2\lambda}C \Sigma )  = \sum_{n=1}^{\infty} \frac{1}{n} \Tr (i\sqrt{2\lambda}C\Sigma)^{n} ; \;  (1-i\sqrt{2\lambda}C \Sigma  )^{-1}C = \sum_{n=1}^{\infty} (i\sqrt{2\lambda}C\Sigma)^{n}C .
\end{eqnarray}
We call the factors $\Tr (i\sqrt{2\lambda}C\Sigma)^{n}$ \emph{loop vertices} \cite{Rivasseau:2007fr} and the factors $(i\sqrt{2\lambda}C\Sigma)^{n} C$ \emph{ciliated vertices} \cite{Gurau:2013pca} or, more simply, \emph{chains}.

We now incorporate the important simplification \eqref{traces1} due to the gauge constraint of our model. 
It ensures that all components of the $\sigma$ matrices factorize trivially from the integral 
\eqref{logi} except the diagonal ones. More precisely since for an intermediate matrix of color $c$
only the $p_c = p'_c $ term contribute, any loop vertex or chain depends only of the \emph{diagonal} part 
$\tau_{c} (p_{c}):=(\sigma_{c})_{p_{c},p_{c}}$
of the six intermediate field matrices previously introduced.  Hence we can reduce the six intermediate matrices in our model 
to six \emph{vector fields} $\tau$ (these diagonal parts)\footnote{This important simplification 
could be interesting for a future constructive analysis of the model.}.
Since each $\tau$ operator is diagonal we conclude also that all propagators occurring in either a $\Tr (i\sqrt{2\lambda}C\Sigma)^{n}$ loop or in a $(i\sqrt{2\lambda}C\Sigma)^{n}C$ chain have the same momentum
$\vec p \in {\mathbb Z}^d$.  Since we remarked that the $\Sigma^c$ 
operators all commute together in the tensor space $\ell^2 ( {\mathbb Z} )^{\otimes 6}$, the value of a loop vertex is a simple sum over the numbers $k_1, \cdots ,k_6$ of insertions
of $\sigma^1, \cdots , \sigma^6$, their total number being $n$. It can therefore
be written as  
\begin{eqnarray}
\Tr{(i\sqrt{2\lambda}C\Sigma)}^{n}&=& [i\sqrt{2\lambda}]^n\sum_{\vec p \in {\mathbb Z}^6, \; \vec k  \in {\mathbb N}^6 \; \vert\; \sum_{c}{k_{c}}=n} \dfrac{n!}{\prod_{c=1}^6 k_c !}
\dfrac{\delta \left( \sum_{c}p_c\right)}{(\vec{p}^2+m^2)^n}\prod_{c=1}^6 \tau_{c} (p_c)^{k_{c}}\nonumber\\
&=& \sum_{\vec p \in \cP  }[ i C_0 (\vec p) \cT(\vec p) ]^n. 
\end{eqnarray}
where we recall that $C_0 (\vec p) =(\vec{p}^2+m^2)^{-1}$, $\cP = \{ \vec p \in {\mathbb Z}^6  \: \vert \;   \sum_{c} p_c  =0 \} $ and 
we define
\bea
\cT(\vec p) =\sqrt{2\lambda} \sum_c  \tau_c (p_c).
\eea
Similarly any chain is a diagonal operator, hence depends on a single momentum $\vec p$  and is non-zero only for $\vec p \in \cP$, with value  
\begin{eqnarray}
(C\Sigma)^n C (\vec p) &=& [i\sqrt{2\lambda}]^n \delta \left( \sum_{c}p_c\right)  \sum_{\vec k  \in {\mathbb N}^6 \; \vert\; \sum_{c}{k_{c}}=n} \dfrac{n!}{\prod_{c=1}^6 k_c !}
\dfrac{1}{(\vec{p}^2+m^2)^{n+1}}\prod_{c=1}^6\tau_{c} (p_c)^{k_{c}}\nonumber\\
&=&  [  i C_0 (\vec p) \cT(\vec p) ]^n  C_0 (\vec p)  . 
\end{eqnarray}

Hence 
\begin{proposition}The partition function of the model is given by
\begin{align}
Z(\bar{J}, J) = \int d\nu (\tau )  e^{ - \sum_{\vec p \in \cP} \bar J (\vec p)   (1- i C_0 (\vec p) \cT(\vec p) )^{-1}  C_0 (\vec p) J  (\vec p)   
-\sum_{\vec p \in \cP} \ln [1- i C_0 (\vec p) \cT(\vec p) ] } .  \label{logitau}
\end{align}
where $d\nu$ is the normalized Gaussian measure on the six vector fields $\tau_c (p)$, each defined on ${\mathbb Z}$,  with trivial covariance
\begin{equation}  \int  d\nu (\tau )  \tau_c (p)  \tau_{c'} (p') = \delta(c,c')  \delta(p,p').  
\end{equation}
\end{proposition}

We want now to describe graphically the Green's functions $G_{2N}$ and the vertex functions of the initial theory in this intermediate field formalism.

\subsection{Graphical Representation}

\begin {figure}[here]
\begin{center}
\includegraphics[scale=0.9]{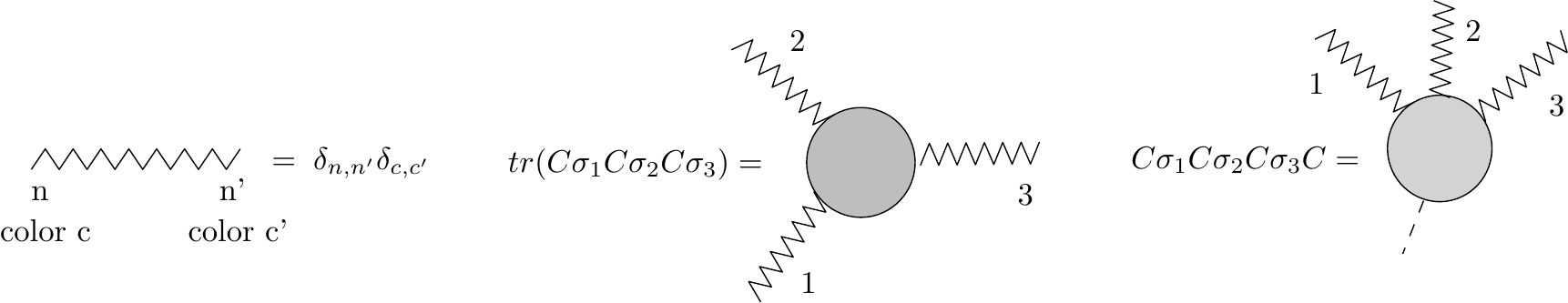}
\caption{Intermediate field graphic representation: propagator, loop vertex and ciliated vertex}
\label{figinterfey0}
\end{center}
\end{figure}

This subsection provides our graphical conventions and Feynman rules for the intermediate field perturbative expansion in the momentum representation.
An intermediated field propagator is represented by a wavy line, which bears a color label $c$ and carries a single momentum $p_c$, 
and correspond to the covariance of a $\tau_c$ intermediate field. 
The loop vertices (which come from deriving the logarithmic interaction in \eqref{logi}), 
are represented by grey disks, to which intermediate field half-lines are attached. 
The chains (which come from deriving the source term in \eqref{logi}) are represented as ciliated lighter gray disks: they are characterized
by a single cilium, represented as a dotted half-line attached to the disk, see Figure \ref{figinterfey0}. A cilium has no color and represents
on its left side the entrance  of the particular momentum $\vec p$ of the chain and on its right side its exit.

\

\begin{figure}[here]
\begin{center}
\includegraphics[scale=0.7]{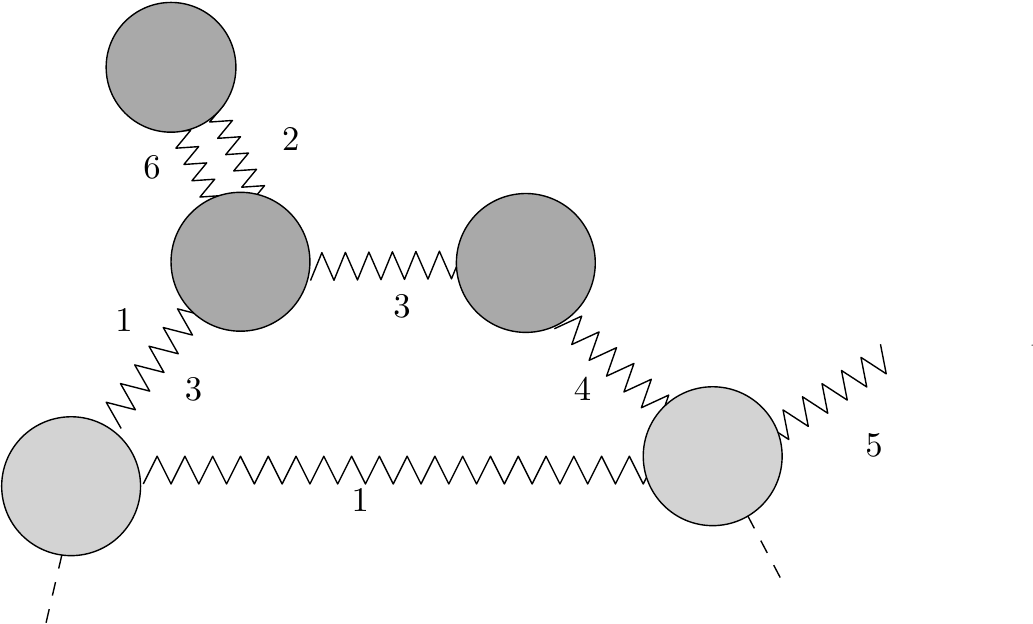} 
\caption{A 4-point graph of the tensorial theory in the intermediate field representation}\label{figinterfey1}
\end{center}
\end{figure}

The former propagators $C$ which were the dotted lines of the initial representation of Figure \ref{figtens}
are now in one-to one correspondence with the \emph{arcs}\footnote{These arcs are often called \emph{corners} 
in the mathematic literature; here we prefer a more physical terminology to convey the fact that arcs are associated to propagators.}
on the boundary of all the disks (both the loop vertices and the ciliated vertices), see Figure \ref{figinterfey1}. 

Green functions $G_{2N}$  of the initial theory can be computed as Feynman graphs with exactly $N$ ciliated vertices and an arbitrary number of loop vertices \cite{Gurau:2013pca}. 
In particular $G_2$ correspond to the sum over connected graphs with exactly one ciliated vertex, and $G_4^c$ to the sum over connected graphs with exactly two ciliated vertices (a generic one is pictured in Figure \ref{figinterfey1}).

\begin{figure}[here]
\begin{center}
\includegraphics[scale=0.7]{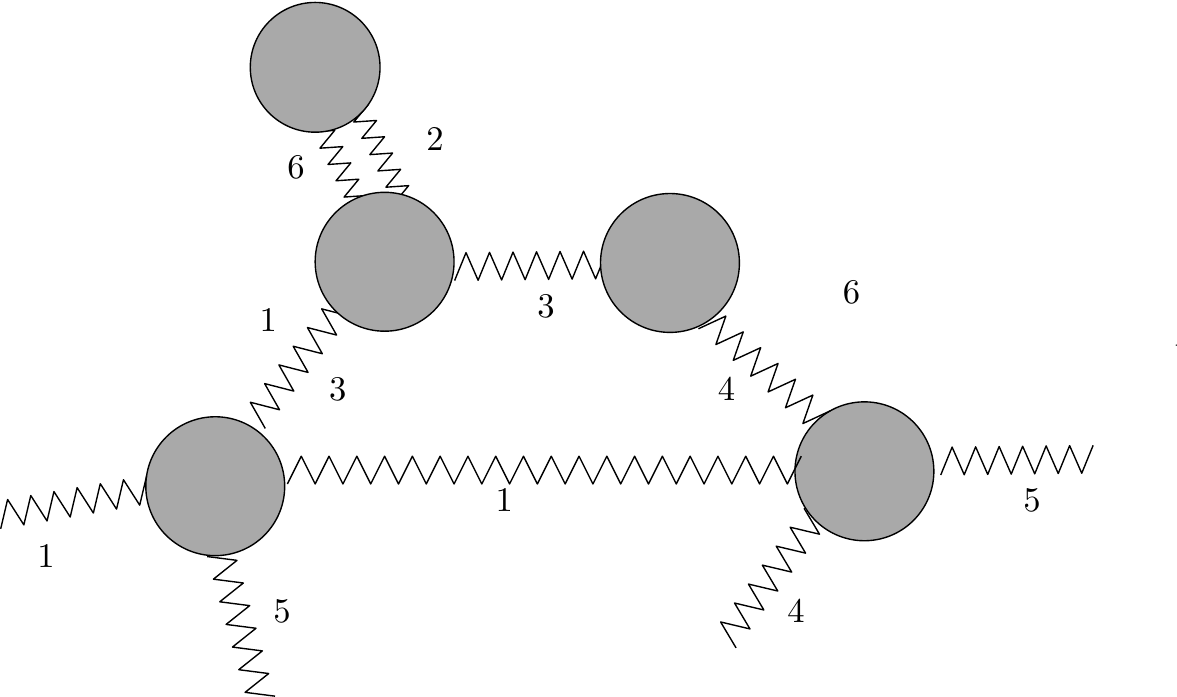} 
\caption{A 4-point graph of the pure intermediate field theory}\label{figinterfey2}
\end{center}
\end{figure}

We can also consider the pure intermediate field theory with the $\bar J$ and $J$ sources put to zero. We introduce new sources $\cJ$ dual to $\tau$. A source $\cJ$ is therefore
a set of six functions $\cJ_c (q_c)$ for $\vec q = \{q_c\} \in {\mathbb Z}^d$.
Introducing the natural notation
$\cJ \cdot \tau  =\sum_{\vec q \in {\mathbb Z}^d}  \sum_{c=1}^d   \cJ_c (q_c) \cdot \tau_c (q_c)$ this pure intermediate field theory 
is defined by the partition function
\begin{equation}
Z(\cJ) = \int d\nu (\tau )  e^{ -  \cJ \cdot \tau 
-\sum_{ \vec p \in \cP} \ln [1- i C_0 (\vec p) \cT(\vec p) ] } .  \label{logipure}
\end{equation}
It has connected Green functions corresponding to 
expectation values of products of $\tau$ fields
\bea
W^{c_1,  \cdots c_N}_{N}(q_1,  \cdots,     q_N  )
=\frac{\partial^N \log Z(\cJ)}   {\partial \cJ_{c_1} (q_1), \
\cdots \partial \cJ_{c_N} (q_N) }\Big|_{\cJ=0}.
\eea
These expectation values are represented by a sum of Feynman
graphs such as those of Figure \ref{figinterfey2}, with a total number of $q$ wavy half-external lines attached to the loop vertices (grey disks), 
each carrying a color $c$ and a single strand momentum $q_c$.

\

By color permutation symmetry, the one point function of the pure theory at color $c$, $W^c_1 (q_c) $ is in fact independent of $c$. 
We call it therefore $W_1 (q) $. It is a function on 
${\mathbb Z}$. Similarly the pure intermediate fields two point function $W^{c_1, c_2}_2 (q_1, q_2)$, which a priori is given as a function of two colors 
$c$ and $c'$, and of two strand momenta $q_1$ and $q_2$, can by color permutation symmetry be described by just two
functions on ${\mathbb Z}^2$, namely $W^{=}_2 (q_1, q_2)$, which corresponds to $c_1 = c_2$, and $W^{\ne}_2 (q_1, q_2)$,
which corresponds to $c_1 \ne c_2$.

\

\

The renormalization of our model will involve only  the \emph{melonic approximation} of the $2$- and 4-point vertex functions $\Gamma^{melo}_{2}$
and $\Gamma^{melo}_{4}$. But there is a simple correspondence
between the \emph{melonic approximation} of the $2N$-point vertex functions $\Gamma^{melo}_{2N}$ of 
the initial theory and the \emph{tree approximation} of the $N$-point  Green functions $W^{tree}_N$ of the pure theory, discovered in 
the context of tensor models \cite{Gurau:2013pca}.
We shall develop it in our case in subsection \ref{subsec3.3} and use it to identify graphically the functions $f$ and $g$ in Theorem \ref{goodth}.
We now return for a while to the initial theory to establish its power counting and renormalization using a multi-scale analysis. This analysis will lead us 
naturally to focus on the melonic approximations which govern renormalization.

\section{Regularization and power counting}
\label{sectionreg}
\subsection{The regularized theory}

Simpler superrenormalizable Abelian TGFT models \cite{Carrozza12013} as well as a just renormalizable non-Abelian model at rank 3 
\cite{Carrozza22013} have been analyzed already using a multiscale expansion. We recall the basic steps of that analysis here, adapting it 
to our specific model. 
	
Like any theory with ultraviolet (UV) divergencies, this model requires a UV cutoff before introducing the renormalization procedure 
(which gives a coherent scheme to extract finite and cut-off independent information). We shall use in this paper both the
parametric cutoffs as in \cite{Carrozza2014} and sharp momentum cutoffs, which are simpler for our model because of its strong momentum conservation rules.

The parametric cutoffs slice the Schwinger parameter. We fix a parameter $ M> 1 $ and define
\begin{eqnarray}
C_{0}(\vec{g},\vec{g'}) &=&\int_{1}^{\infty} \mathrm{d}\alpha e^{-\alpha m^{2}} \int{dh \prod_{c=1}^{d} K_{\alpha}(g_{c}hg{'}_{c}^{-1})}
\nonumber\\
C_{i}(\vec{g},\vec{g'})&=&\int_{M^{-2i}}^{M^{-2(i-1)}} \mathrm{d}\alpha e^{-\alpha m^{2}} \int{dh \prod_{c=1}^{d} K_{\alpha}(g_{c}hg{'}_{c}^{-1})}, i \neq 0.
\end{eqnarray}
We choose the UV-regulator $\Lambda$ so that $\Lambda=M^{-2\rho}$, and the complete propagator $C_{\Lambda}\equiv C^{\rho}$ is then given by:  
\begin{equation}
C^{\rho}=\sum_{i=0}^\rho C_i.
\end{equation}

A corresponding sharp momentum cutoff $\chi_{\le\rho} (\vec p) $ 
is 1 if $\vert \vec p\vert^2  \le M^{2\rho}$ and zero otherwise. The theory with cutoff $\rho$ is defined by using the covariance 
\begin{equation}
C^{\rho}(\vec p) =C(\vec p)  \chi_{\le\rho} (\vec p) .
\end{equation}
Then we slice the theory according to 
\begin{equation}
C^{\rho}(\vec p) =\sum_{i=1}^\rho C_i (\vec p),\; C_i (\vec p) = C(\vec p) \chi_i (\vert \vec p\vert^2) \label{cutoffsha}
\end{equation}
where $\chi_1$  is 1 if $ \vert \vec p\vert^2  \le M^{2}$ and zero otherwise and for $i\ge 2$ $\chi_i$ is 1 if $M^{2(i-1)} < \vert \vec p\vert^2  \le M^{2i}$ and zero otherwise.

A subgraph $S \subset G$ in an initial Feynman graph is a certain subset of lines (propagators $C$) plus the vertices attached to them; the half-lines attached to the vertices of $S$ (whether external lines of $G$ or half-internal lines of $G$ which do not belong to $S$) form the external lines of $G$. Translating
to the intermediate representation, we find that a subgraph should be a \emph{set of arcs} of the intermediate field representation, plus all
the wavy edges attached to these arcs. The external lines are then the (half)-arcs attached to these wavy edges which do not belong to $S$.

A \emph{vertex} of the initial representation is called external for $S$ if it is hooked to at least one external line for $S$. Similarly a wavy \emph{line} of the intermediate representation will be called external to $S$ if it hooks to at least one external \emph{arc}.

Particularly interesting subgraphs in the intermediate field representation are those for which the set of arcs are exactly those of a set $S \subset \cL\cV$ of loop vertices (excluding any chain, so no arc belongs to any ciliated vertex). Let us call such subgraphs \emph{proper intermediate} or PI.  
Remark that any PI graph is automatically 1PI in the initial representation (since all arcs belong to at least one loop, the one of their loop vertex). Also
any PI graph can be considered amputated, hence as a graph for a particular vertex function. The converse is not true and many graphs for vertex functions
do not correspond to PI graphs in the intermediate representation.

PI subgraphs can be represented as graphs of the pure intermediate theory, simply by omitting the two half-arcs at the end of each external wavy line.
In our model their amplitude depends only of the single strand momentum entering the wavy line, not of the full momentum of the two half-arcs hooked at its end.

We shall see that in our theory only very particular non-vacuum connected
subgraphs are superficially divergent, namely PI graphs which are trees with at most two external lines.

\subsection{Multiscale analysis}

The multi-scale analysis \cite{Rivasseau:1991ub} allows to renormalize in successive steps, in the Wilsonian spirit.
It attributes a scale to each line $\ell \in \mathcal{L} (\mathcal {G}) $ of any amplitude
of any Feynman graph $\mathcal {G}$. 

Let us start by establishing multi-scale power counting. We can perform this analysis both with parametric or sharp cutoffs, ending with the same conclusions.
In this subsection we use the sharp cutoffs since they attribute
the same scale to all arcs of any loop vertex or chain, hence a single scale to any loop vertex of the intermediate field representation.

The amplitude of a graph $ \mathcal {G} $, $ \mathcal {A} (\mathcal {G}) $, with fixed external momenta, is thus divided into the sum of all the scale attributions
$ \mu = \{i_{\ell}, \ell \in \cL (  \mathcal {G} ) \} $, where $i_\ell$ is the scale of the momentum $p$ of line $\ell$:
\begin{equation}
\mathcal{A}(\mathcal{G})=\sum_{\mu}\mathcal{A}_{\mu}(\mathcal{G}).
\end{equation}
At fixed scale attribution $\mu$, we can identify the power counting in powers of $M$. The essential role is played by the 
subgraph $\mathcal{G}_i$ formed by the subset of lines of $\mathcal{G}$ with scales higher than $i$. By the momentum conservation rule
along any loop vertex, this subgraph is automatically a PI subgraph
which decomposes into $k(i)$ connected PI components : $\mathcal{G}_i=\cup_{k=1}^{k(i)} \mathcal{G}_i^{(k)}$. These connected components
form when $(i,k)$ take all possible values
an abstract tree for the inclusion relation (the famous Gallavotti-Nicol\`o tree \cite{GalNic}). We have 
\begin{theorem}
The amplitude $\mathcal{A}_{\mu}(\mathcal{G})$ 
is bounded by:
\begin{equation}
\quad|\mathcal{A}_{\mu}(\mathcal{G})|\leqslant K^{L({\mathcal{G}})} \prod_{i}\prod_{k=1}^{k(i)}M^{\omega(\mathcal{G}^k_i)} , \ K>0 ,
\end{equation}
and the divergence degree $\omega(\mathcal{H})$ 
of a connected subgraph $\mathcal{H}$ is given by:
\begin{equation}
\omega(\mathcal{H})=-2L(\mathcal{H})+F(\mathcal{H})-R(\mathcal{H}),
\end{equation}        
where $L(\mathcal{H})$ and $F(\mathcal{H})$ are respectively the number of lines and internal faces of the subgraph $\mathcal{H}$, and $R(\mathcal{H})$ is the rank of the adjacency matrix $\epsilon_{\ell f}$ for the lines and faces of $\mathcal{H}$.
\end{theorem}
\textit{Proof}:
Obviously we have (for $K=M^2$)
\begin{equation}
\vert C_i(\vec{p}) \vert \leq K \delta (\sum_c p_c)  M^{-2i}  \chi_{\le i } ( \vec p).
\end{equation}
Fixing the external momenta of all external faces
the Feynman amplitude (in this momentum representation) is bounded by
\begin{align}
|A_{\mu}(\mathcal{G})|\leq & \left[ \prod_{\ell\in\mathcal{L}(\mathcal{G})}KM^{-2i_\ell}\right]  
\prod_{f\in F_{int}(\mathcal{G})}  \sum_{p_f\in {\mathbb Z}}  \prod_{\ell \in \partial f} 
\chi_{\le i_\ell } ( \vec p)  \prod_{\ell\in\mathcal{L}(\mathcal{G})} \delta (\sum_c p^\ell_c ) . 
\end{align}
The key to multiscale power counting is to attribute the powers of $M$ to the $\mathcal{G}_i^{(k)}$ connected components.
For this, we note that, trivially: $M^i=M^{-1}\prod_{j=0}^i M$, a trivial but useful identity which allows e.g. to rewrite 
$\prod_{\ell\in L(G)}M^{-2i_l}=M^2\prod_{\ell\in L(G)}\prod_{i=0}^{i_\ell}M^{-2}$. Then, inverting the order of the double product leads to
\begin{equation}
\prod_{\ell\in L(\mathcal{G})}M^{-2i_\ell}=\prod_i\prod_{\ell\in \mathcal{L}(\cup_{k=1}^{k(i)}\mathcal{G}_i^k)}M^{-2}=\prod_i\prod_{k=1}^{k(i)}\prod_{l\in \mathcal{L}(\mathcal{G}_i^k)}M^{-2}=\prod_i\prod_{k=1}^{k(i)}M^{-2L(\mathcal{G}_i^k)}.
\end{equation}
The goal is now to optimize the cost of the sum over the momenta $p_f$ of the internal faces. 
Summing over $p_f$ with a factor $\chi_{\le i } ( \vec p)$ leads to a factor $KM^i$, hence we should sum with the smallest values $i(f)$ of slices $i$ for the lines $\ell \in \partial f$
along the face $f$. This is exactly
the value at which, starting form $i$ large and going down towards $i=0$ the face becomes first internal for some $\mathcal{G}_i^k$.
Hence in this way we could bound the sums $\prod_{f\in F_{int}(\mathcal{G})}  \sum_{p_f\in {\mathbb Z}} $ by
\begin{equation}
\prod_i\prod_{k=1}^{k(i)}M^{F(\mathcal{G}_i^k)}.
\end{equation}
However this can be still improved, because we have not yet taken into account the gauge factor $\prod_{\ell\in\mathcal{L}(\mathcal{G})} \delta (\sum_c p _c ) $. 
It clearly tells us that some sums over $p_f$ do not occur at all. How many obviously depends of the \emph{rank} $R$ of the incidence matrix $\epsilon_{\ell f}$.
Indeed rewriting the delta functions in terms of the $p_{f(\ell, c)}$ we have
\begin{equation}
\prod_{\ell\in\mathcal{L}(\mathcal{G})} \delta (\sum_c p^\ell_c )= \prod_{\ell\in\mathcal{L}(\mathcal{G})} \delta (\sum _f \epsilon_{\ell f} p_f) .
\end{equation}
Hence writing the linear system of $L$ equations $\sum _f \epsilon_{\ell f} p_f =0$ corresponding to the delta functions
we can solve for $R$ momenta $p_f$ in terms of $L-R$ others. It means that in the previous argument we should pay only for $F-R$ sums over
internal face momenta instead of $F$
\footnote{Remark that the remaining product unused or redundant $\delta$ functions are simply bounded by 1 because the $p_f$ variables are discrete,
hence the $\delta$ function are simply Kronecker  symbols, all bounded by 1; of course this would not be true for continuous variables as a product of redundant
$\delta$ distributions in the continuum is ill-defined.}. 

This argument can be made more precise and rigorous and distributed over all scales 
starting from the leaves of the Gallavotti-Nicol\`o tree (the smallest subgraphs $\mathcal{G}_i^k$) and progressing towards the root we can select faces
such that the restricted sub-matrix $\epsilon_{\ell f}  $ still has maximal rank $R(\mathcal{G}_i^k)$ in each $\mathcal{G}_i^k$. We 
discard the other faces decay factor. Then we can select lines so as to find a restricted \emph{square} submatrix $\epsilon_{\ell f}  $ with maximal rank $R(\mathcal{G}_i^k)$ in each 
$\mathcal{G}_i^k$. This leads to
\begin{equation}
|A_{\mu}(\mathcal{G})| \leq K^{\mathcal{L}(\mathcal{G})}\prod_i\prod_{k=1}^{k(i)}M^{-2L(\mathcal{G}_i^k)+F(G_i^k)-R(\mathcal{G}_i^k)} = K^{|L(G)|}\prod_i\prod_{k=1}^{k(i)}M^{\omega(\mathcal{G}_i^k)}.   \label{powerbound}
\end{equation}
This equation completes the proof, and the exponent $\omega(\mathcal{G}_i^k)=-2L(\mathcal{G}_i^k)+F(\mathcal{G}_i^k)-R(\mathcal{G}_i^k)$ identifies the divergence degree.
\qed

\subsection{Melonic graphs}
\label{subsec3.3}

In this subsection, we will determine the nature of PI superficially divergent graphs, which are those with positive divergence degree $\omega \ge 0$. 
We shall establish that they are \emph{melonic} \cite{review}
in the ordinary representation, and \emph{trees}
in the intermediate field representation.

Consider first the case of a PI vacuum subgraph. If it is a tree on $n$ loop vertices, it has $L=2(n-1)$ arcs, $5n +1 $ faces (since each wavy line glues two faces) and it is easy to check by induction (adding leaves one by one from a root) that the rank $R$ of the $\epsilon$ matrix is maximal, namely $n$. 
Hence $\omega = -4(n-1) + 5n +1 -n =5$ in this case.

Next let us consider the case of a PI tree subgraph with $N$ external wavy lines, hence $2N$ external arcs. 

\begin{itemize}
\item if $N=1$ the subgraph is a two point function and the single external wavy line adds one arc, suppresses one face and does not change the rank, hence $\omega = 2$ in this case.

\item if $N=2$ the subgraph is a four point function and the two external wavy lines adds two arcs. If they have different colors, or have the same color $c$
and hook to two components of the tree not connected by lines of color $c$, then they open two different faces and do not change the rank, so that 
$\omega = -1$. However there is a special case,
when the two external wavy lines have same color and hook to the same loop vertex or to different loop vertices joined by a 
path in the tree made of wavy line all of the same color $c$. In that case and only that case,  the wavy lines open only the single face of color $c$ 
common to all loop vertices along this path, the rank again has not changed and $\omega = 0$.

\item if $N>2$, each new external line takes $L$ into $L+1$, can either keep $F$ unchanged (if it hits an already open face), in which case $R$
is also unchanged, or takes $F$ to $F-1$, in which case either $R$ is unchanged or goes to $R-1$; Hence $\omega$ decreases at least by 1.
This proves
\begin{equation}  \omega(\cG)  \le  - (N-2) \;\;{\rm if} \;\; N >2. \label{extlegs}
\end{equation}
\end{itemize}

Consider next the case of a PI vacuum subgraph with $N$ external wavy lines and $q$ wavy loops. 
We can first pick a tree of wavy lines then add the wavy loops one by one. Each 
added loop creates two new arcs and changes the number of faces by -1, 0 or 1. It can change the rank at most by 1, and when it creates a face, then the rank cannot decrease (the matrix $\epsilon$ becoming bigger). Hence 
\begin{equation}  \omega(\cG)  \le  - (N-2) - 3q \;\;{\rm if} \;\; N >2. \label{extlegsloops}
\end{equation} 
In particular if $N=1$ and $q\ge 1$ we have $\omega(\cG)  \le -1$ and the graph is convergent.

Finally it remains to study the case of non-vacuum, non-PI graph. Since they add at least one new arc to a PI graph,
it is easy to check they have $\omega <0$, except in two particular cases corresponding both to one-particle reducible graphs:

\begin{itemize}
\item a chain of arcs joining PI two-point trees, with one of them at both ends. Such subgraphs are one-particle reducible two point subgraphs of the initial theory with $\omega =2$.

\item a chain of arcs joining PI two-point trees, with one of them at a single of its two ends. Such subgraphs are one-particle reducible four point subgraphs 
of the initial theory, with $\omega =0$.
\end{itemize}

These cases are not interesting since such subgraphs cannot occur as $\mathcal{G}_i^k$s and, as is well known, renormalization can be restricted
to IPI subgraphs.

These results in particular show that the degree of divergence $\omega$ does not depend on the number of vertices, 
but only on the number of external lines. This is typical of a just renormalizable field theory.

Trees in the intermediate representation correspond to melonic subgraphs in the ordinary representation \cite{Gurau:2013pca}.
Hence we have proved, in agreement with the other renormalizable TGFT's:

\begin{theorem} The only superficially divergent PI subgraphs are melonic in the ordinary representation, with two or four external ordinary lines. 
In the intermediate representation, amputating the trivial 
external arcs, they are PI trees with a single external wavy line, 
or with two external wavy lines of the same color carrying the same strand momentum.   \label{theordivequalmelo}
\end{theorem} 

Melonic graphs are graphs with zero \emph{degree}\footnote {The degree in question is the "degree of the colored graph", which characterizes the dominant order of the large-N limit of tensor models. It should not be confused with the degree of divergence, and we denote it by $\varpi(\mathcal{G}_c)$.}, hence for which all \emph{jackets} are planar. 
We include for completeness brief definitions of these two notions, referring to \cite{review} for more details.

\begin{definition}[Jackets] A \textit{jacket} $\mathcal{J}$ of a regular $d+1$ colored graph $\mathcal{G}_{c}$ is the canonical ribbon graph
associated to $\mathcal{G}_{c}$ and to a (D+1)-cycle $\xi$ up to orientation. It has the same number of lines and vertices than $\mathcal{G}_{c}$, 
but contains only a subset of the faces, those with consecutive colors in the cycle $\mathcal{F}_{\mathcal{J}}=\left\lbrace f\in \mathcal{F}_{\mathcal{G}_{c}}|f=({\xi}^q(0),{\xi}^{q+1}(0)), q\in \mathbb{Z}_{D+1}\right\rbrace $.
\end{definition}

Hence there are $d!/2$ jackets at rank $d$ and to each jacket is associated a Riemann surface of genus
$g_{\mathcal{J}}$. 
\begin{definition}[Degree] The degree $\varpi(\mathcal{G}_{c})$ is by definition the sum over the \emph{genus} of all the jackets:
\begin{equation}
\varpi(\mathcal{G}_{c})=\sum_{\mathcal{J}}g_{\mathcal{J}} \Rightarrow \varpi(\mathcal{G}_{c})\geq 0 .
\end{equation}
\end{definition}
The degree governs the $1/N$ tensorial expansion since the number of faces is a monotonically decreasing function of the degree. 
Melonic graphs have maximal number of faces at a given perturbation order.
More precisely
\begin{lemma}
The number of faces $F_c$ of
$\mathcal{G}_{c}$ is related to the number of black vertices $p$ and to the dimension $d$ by:
\begin{equation}
F_c=\dfrac{d(d-1)}{2}p+d-\dfrac{2}{(d-1)!}\varpi(\mathcal{G}_{c}) . \label{numberfaces}
\end{equation}
\end{lemma}

A tensorial graph $\cG$ having a unique colored extension $\cG_c$, we can extend the notion of degree to tensorial graph. Since the colored extensions of type 1 vertices 
of our theory all have the same number of inner faces (faces without color 0), the degree of $\cG_c$ again governs the number of faces of $\cG$, which are the bicolored faces
of $\cG_c$ which includes color 0. In our case the vertices of $\cG$ all have 25 inner faces and $p=2$ black vertices, so that \eqref{numberfaces} tells us
\begin{equation}
F(\cG) =5V +6-\dfrac{1}{60}\varpi(\mathcal{G}_{c}) . \label{numberfaces1}
\end{equation}

Returning to Theorem \ref{theordivequalmelo} we can precise the divergent part of the theory in the language of the previous section.
$\Gamma_2^{melo}$ and $\Gamma_4^{melo}$ are naturally defined as the melonic approximations to $\Gamma_2$ and $\Gamma_4$
and Theorem \ref{theordivequalmelo} indeed  proves that $\Gamma_2- \Gamma^{melo}_2$ and $\Gamma_{4,mono}- \Gamma^{melo}_{4,mono}$
are superficially convergent. Moreover they express simply as tree approximations of the 
pure $\tau$ intermediate field theory: we have 
\begin{equation}
\Gamma_2^{melo} (\vec p) = \sqrt{2 \lambda} \sum_{c=1}^6   W^{tree}_1 (p_c),\; 
\Gamma_4^{melo} (p_c,p_c') = \sqrt{2 \lambda} \delta(p_c, p'_c) W^{=,tree}_{2} (p_c, p_c) \label{goodeq1}
\end{equation}
where $W^{tree}_1 $ and $W^{=,tree}_{2}$ are respectively the tree approximation to $W_1 $ and $W^{=}_{2}$.

But Theorem \ref{theordivequalmelo} contains still an additional information on the divergent part of $\Gamma_4^{melo}$.
Defining  $W^{=,tree}_{2,mono}$ as the part of $W^{=,tree}_{2}$ in which
all wavy lines along the unique path joining the two external lines must be of the same color $c$ than these two external lines,
it states that the difference  $W^{=,tree}_{2}- W^{=,tree}_{2,mono}$ is also ultraviolet finite, hence can be neglected in the following
section on renormalization.

Since \eqref{goodeq1} is nothing but \eqref{goodeq} with $f= -i \sqrt{2/ \lambda} W^{tree,c}_1$ and $g= W^{tree}_{2,mono}$, 
this completes the proof of Theorem \ref{goodth}.

\subsection{Uniform Convergent Bounds}

An important aspect of the multiscale analysis is that it provides easily a uniform exponential bound on \emph{convergent} amplitudes: 
\begin{theorem}
\textbf{(Uniform Weinberg theorem)} The amplitude $A(\mathcal{G})$ for a completely convergent connected graph 
$\mathcal{G}$ (i.e. a graph for which $\omega(\mathcal{H})<0 \,\forall \mathcal{H} \subset \mathcal{G}$) is uniformly bounded in terms of its size, 
i.e. there exists a constant $K$ such that if $n$ is the order (number of vertices)
of the graph: 
\begin{equation} | \cA (\mathcal{G})|\leq K^{n(\mathcal{G})}. \label{equwein}
\end{equation}\label{theowein}
\end{theorem}

\prf We assume $N(\cG) \ge 1$, so that $\forall \mathcal{H} \subset \mathcal{G}, N(\cH) \ge 1$ (the vacuum case $N(\cG) =0$ is an easy extension
left to the reader).
\eqref{extlegs} implies that for a convergent PI graph with $2N>4$ external arcs
\begin{equation}
\omega (\cH) \le - N(\cH)/3 = - 2N(\cH)/6.
\end{equation}
This is also true if $\cH$ is convergent with $N = 1$ or 2, since we saw that in this case $\omega \le -1 \le - N(\cH)/3$.
For a $\phi^4$ graph of order $V=n$ with $2N$ external legs, we have $2L = 4V +2N$. Therefore \eqref{powerbound} implies
that for another constant $K$
\begin{equation}
\cA(\mathcal{G}) \le K^n \sum_\mu \prod_i\prod_{k=1}^{k(i)}M^{-2N(\mathcal{G}_i^k)/6 }.
\end{equation}
Let us now define
\begin{equation}
i_v(\mu)=\sup_{\ell \in L_v(\mathcal{G})}i_\ell (\mu) , e_v(\mu)= \inf_{ \ell \in L_b(\mathcal{G})}i_\ell(\mu),
\end{equation}
where $v$ denotes a vertex $v\in\mathcal{G}$, and $L_v(\mathcal{G})$ the set of its external (half)-lines. 
$v$ is external to a high subgraph $\mathcal{G}_i^k$  if and only if $e_b < i \leq i_b$, and then it is hooked to at least one of the $2N(\mathcal{G}_i^k)$ external half-lines of $\mathcal{G}_i^k$. Therefore
\begin{equation}
\prod_{i,k}M^{-2N(\mathcal{G}^{(k)}_i)/6} \leq \prod_{i,k}\prod_{v\in \mathcal{G}^{(k)}_i  | e_v<i\leq i_v}M^{-1/6}.
\end{equation}
Using the fact that there are at most $4$ half-lines, and thus $6=4\times 3/2$ pairs of half-lines hooked to a given vertex, and that, for two 
external lines $\ell$ and $\ell'$ of a vertex $v$, $| e_v-i_v| \geq |i_\ell -i_{\ell'}|$, we obtain:
\begin{equation}
\cA(\mathcal{G}) \le K^n \sum_\mu \prod_{v} \prod_{(\ell,\ell') \perp v }M^{-\frac{|i_\ell-i_{\ell'}|}{36}}, \label{vertibou}
\end{equation}
where the product over $(\ell,\ell') \perp v$ means the product over all pairs of half-lines hooked to $v$. 
The bound means that there is exponential decay in scale differences between all such pairs{\footnote{\eqref{vertibou} is of course
a very sloppy estimate, that could be easily improved. For instance we could take advantage of the momentum representation conservation rules to remark that 
only one pair of different scales is in fact hooked to any vertex, rather than 6, but it won't change the structure of the result, only improve numerical constants.}.
Organizing the sum over $\mu=\{i_{\ell}\}$ 
along a tree of lines of $\cG$ as in \cite{Rivasseau:1991ub}, 
it is easy to bound it by $K^{L(\cG}$, hence to complete the proof 
of \eqref{equwein}, hence of Theorem \ref{theowein}. \qed

The next section is devoted to renormalization of the model and to a computation of its beta function.

\section{Perturbative Renormalization and Flow}
\label{sectionren}
	
Renormalization consists, after having identified the ``dangerous" subgraphs $\mathcal{G}_i^k$ (those with $\omega\ge 0$), 
in subtracting from them their local Taylor approximation (the ``counter-terms"), up to cancelation of the divergencies, hence up to order $\omega$. 
Then one should compute how renormalization changes the interaction from bare to renormalized, hence
compute the flow of the theory from the ultraviolet to the infrared regime.

\subsection{Perturbative renormalization and counter-terms}

Our goal in this section is to check that, as stated in \cite{Carrozza2014,Samary:2013xla}
 
\begin{theorem} The $U(1)$ model with $T^4$ interaction at rank 6 is just renormalizable and asymptotically free.
\end{theorem}
	  
The perturbative renormalization implies the following redefinitions
\begin{equation}
\phi=Z^{1/2}(\Lambda)\phi_{r}, \; \bar \phi=Z^{1/2}(\Lambda)\bar \phi_{r},
\end{equation}
\begin{equation}
\lambda=Z^{-2}(\Lambda)Z^{1/2}_{\lambda}(\Lambda)\lambda_{r}=\mathcal{Z}_{\lambda}^{1/2}\lambda_{r},
\end{equation}
\begin{equation}
m=Z^{-1/2}(\Lambda)Z^{1/2}_{m}(\Lambda)m_{r}=\mathcal{Z}_{m}^{1/2}m_{r},
\end{equation}
and the UV-regularized generating partition function is:
\begin{align}
\mathcal{Z}:&=\int d\mu_{C(Z^{-1/2}Z^{1/2}_{m}m_r)}(Z^{1/2}\phi_r,Z^{1/2}\bar{\phi}_r)e^{Z^{1/2}_{\lambda}\lambda_{r}\sum_{i=1}^{6}{ Tr_{b_{i}}(\bar{\phi}_r \phi_r)}}\nonumber\\
&=\int d\mu_{C(Z^{-1/2}Z^{1/2}_{m}m_r)/Z}(\phi_r,\bar{\phi}_r)e^{Z^{1/2}_{\lambda}\lambda_{r}\sum_{i=1}^{6}{ Tr_{b_{i}}(\bar{\phi}_r \phi_r)}}.
\end{align}

In these definitions, the ``r" subscript applies to the renormalized quantities. 
The mass and wave function counter terms can be absorbed in the covariance
\begin{align}
\int d\mu_{C(Z^{-1/2}Z^{1/2}_{m}m_r)/Z}(\phi_r,\bar{\phi}_r)\phi_r(\vec{\theta}) \bar{\phi}_r(\vec{\theta '})&=\sum_{\vec{p}}\dfrac{1}{Z}\dfrac{\delta(\sum_c p_c)}{\vec{p}^2+Z^{-1}Z_m m^2_{r}}e^{i\vec{p}\cdot(\vec{\theta}-\vec{\theta}')}\nonumber\\
&=\sum_{\vec{p}}\dfrac{\delta(\sum_c p_c)}{\vec{p}^2+m^2_{r}}\dfrac{1}{1+\dfrac{\delta_Z\vec{p}^2+\delta_{m^2} m^2_{r}}{\vec{p}^2+m^2_{r}}
}e^{i\vec{p}\cdot(\vec{\theta}-\vec{\theta}')},
\end{align}
with $\delta_Z = Z -1$, $ \delta_{m^2}  = Z_m -1$.
Identifying this covariance with the one of the initial bare theory means that the (bare) propagator of the theory rewrites in terms of renormalized quantities as
\begin{align} C =
\dfrac{\delta(\sum_c p_c)}{\vec{p}^2+m^2_{r}
+\delta_Z\vec{p}^2+\delta_{m^2} m^2_{r}}.
\label{equeff2pt}
\end{align}

As well known the renormalized parameters in a BPHZ scheme are obtained in terms of the bare ones through the \emph{vertex functions}, 
which are the one-particle irreducible amputated 
functions. In our model the power counting analysis of the previous section showed that we need only to renormalize
the 2 point vertex function $\Gamma_2$ (self-energy),
and the four point vertex function $\Gamma_4$.

\subsection{The renormalization group flow}

The basic idea of the renormalization group is the following. All correlation functions are invariant under an infinitesimal dilatation $ s:= 1 + \delta $ of the 
ultraviolet cut-off $ \Lambda $ with a simultaneous redefinition of the coupling constants, mass, and field normalization:
\begin{align}
\Lambda \rightarrow \Lambda(1+\delta), \;\; m \rightarrow m+\delta m , \;\;
\lambda \rightarrow \lambda+\delta \lambda , \;\;
Z \rightarrow Z(1+\delta Z) .
\end{align}

Renormalized quantities parametrize a given trajectory of the RG flow. We have the relations:
\begin{align}
\phi_{s\Lambda}=Z^{1/2}(s\Lambda)\phi_{r}, \;\;
m(s\Lambda)=\mathcal{Z}_m^{1/2}(s\Lambda)m_{r}, \;\;
\lambda(s\Lambda)=\mathcal{Z}_{\lambda}^{1/2}(s\Lambda)\lambda_{r},
\end{align}
involving
\begin{align}
\phi_{s\Lambda}=Z^{1/2}(s\Lambda)Z^{-1/2}(\Lambda)\phi_{\Lambda}=:Z^{1/2}(s)\phi_{\Lambda},\\
m(s\Lambda)=\mathcal{Z}_m^{1/2}(s\Lambda)\mathcal{Z}_m^{-1/2}(\Lambda)m(\Lambda)=:\mathcal{Z}_m^{1/2}(s)m(\Lambda), \\
\lambda(s\Lambda)=\mathcal{Z}_{\lambda}^{1/2}(s\Lambda)\mathcal{Z}_{\lambda}^{-1/2}(\Lambda)\lambda(\Lambda)=:\mathcal{Z}_{\lambda}^{1/2}(s)\lambda(\Lambda).
\end{align}

These relations give the transformations of field, mass and couplings of two theories with different cut-offs, hence along the same trajectory of the RG flow. 
They imply trivially the invariance of the renormalized correlation functions along a given trajectory. This invariance translates into a differential equation for the correlation functions describing the evolution of the RG flow, namely the so called Callan-Symanzik (CS) equation. Writing
\begin{equation}
G_{\Lambda,m_{\Lambda},\lambda_{\Lambda}}^{2N}(\{\vec{\theta}_i\})=Z^{N}(\Lambda)G_{r,m_r,\lambda_r}^{2N}(\{\vec{\theta}_i\}),
\end{equation}
with
\begin{equation}
[Z(s\Lambda)Z^{-1}(\Lambda)]^{N}G_{\Lambda,m_{\Lambda},\lambda_{\Lambda}}^{2N}=G_{s\Lambda,m_{s\Lambda},\lambda_{s\Lambda}}^{2N}.
\end{equation}
and developing, with $ s:= 1 + \delta $, we get
\begin{equation}
Z(s\Lambda)Z^{-1}(\Lambda)=\left( Z(\Lambda)+\Lambda\dfrac{dZ}{d\Lambda}\delta\right) Z^{-1}(\Lambda)=1+\delta\Lambda\dfrac{d}{d\Lambda} \ln{Z},
\end{equation}
\begin{equation}
G_{s\Lambda,m_{s\Lambda},\lambda_{s\Lambda}}^{2N}=G_{\Lambda,m_{\Lambda},\lambda_{\Lambda}}^{2N}+\Lambda\delta \left\lbrace \dfrac{\partial}{\partial\Lambda}+\dfrac{d\lambda}{d\Lambda}\dfrac{\partial}{\partial \lambda}+\dfrac{dm^2}{d\Lambda}\dfrac{\partial}{\partial m^2}\right\rbrace G_{\Lambda,m_{\Lambda},\lambda_{\Lambda}}^{2N}.
\end{equation}
Gluing the pieces, we obtain the CS equation:
\begin{equation}
\left\lbrace  \Lambda \dfrac{\partial}{\partial\Lambda}+\beta(\lambda)\dfrac{\partial}{\partial \lambda}+m^2 \gamma_{m^2}(\lambda) \dfrac{\partial}{\partial m^2}+N \gamma (\lambda) \right\rbrace G_{\Lambda,m_{\Lambda},\lambda_{\Lambda}}^{2N}=0, \label{callansym}
\end{equation}
with the following definitions:
\begin{equation}
G_{\Lambda,m_{\Lambda},\lambda_{\Lambda}}^{2N}:=\dfrac{1}{\mathcal{Z}}\int d\mu_{C_{\Lambda}}(\bar{\phi},\phi)\prod_{j=1}^{N}\bar{\phi}^{(j)}(\vec{\theta}_j)\phi^{(j)}(\vec{\theta}'_j)e^{-S_{int}(\bar{\phi},\phi)},
\end{equation}
\begin{equation}
\beta:=\Lambda\dfrac{d\lambda}{d\Lambda}, \; \gamma:=-\Lambda\dfrac{d}{d\Lambda} \ln{Z },\;  
\gamma_{m^2}:=\Lambda\dfrac{d}{d\Lambda} \ln{m^2}.  \label{betafunction}
\end{equation}
We analyze now this equation at first order (one loop).

\subsection{One loop self energy}

We start by computing the corrections to the propagator. At one loop, the only melonic (hence divergent) 
graph is  pictured in Figure \ref{renmaa}. 

\

\begin{figure}[here]
\begin{center}
\includegraphics[scale=0.7]{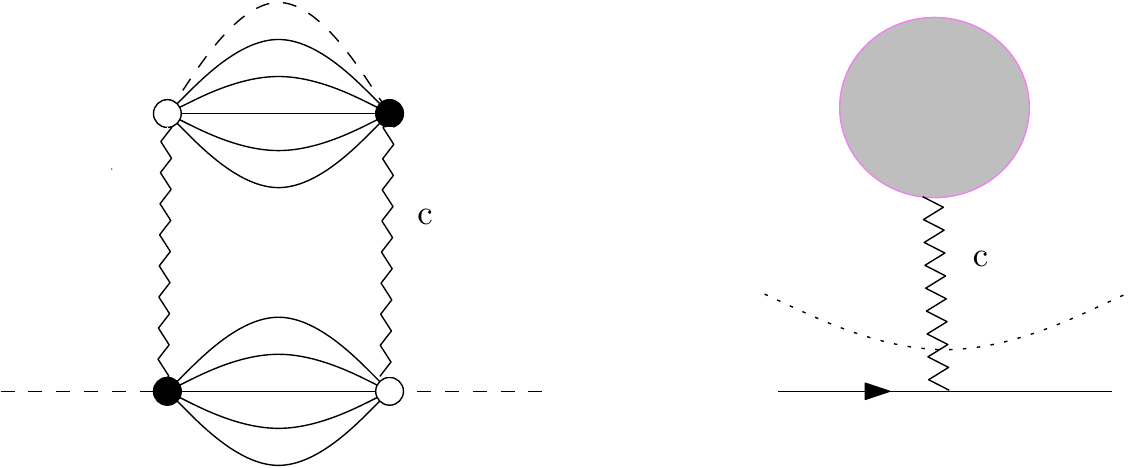}
\caption{The self-energy contribution at one-loop} \label{renmaa}
\end{center}
\end{figure}

Its value is
\begin{equation}
\Gamma_2^{melo,1}(\vec{p})=-\sum_c \sum_{q_{c', c' \ne c} } 2\lambda \dfrac{\delta (\sum_{c'} q^c_{c'})}{(\vec{q}^c)^2+m^2}=-2\lambda \sum_c\sum_{q^c_{c'}\, c'\neq c} \dfrac{\delta (\sum_{c'} 
q^c_{c'})}{(\vec{q}^c)^2+m^2},
\end{equation}
where $\vec{q}^c \in {\mathbb Z}^d$ has components $q^c_{c'}= (q_1,q_2,...,q_c=p,...,q_d)$.

Let's use the Schwinger representation to rewrite the denominator as an integral of an exponential, with 
UV cutoff on the parameter $\alpha$. It gives
\begin{align}
\Gamma_2^{melo,1}(\vec{p})&=-2\lambda \sum_c \int_{1/\Lambda^2}^{\infty} d\alpha e^{-\alpha m^2} \sum_{q^c_{c' \neq c}}  \delta (\sum_{c'} q^c_{c'}) e^{-\alpha (\vec{q}^c)2}
\nonumber\\
&=-2\lambda \sum_c\int_{1/\Lambda^2}^{\infty} d\alpha e^{-\alpha m^2} \int_{0}^{2\pi} \dfrac{d\beta}{2\pi} \sum_{q^c_{c' \neq c}} e^{i\beta\left( \sum_{c'} q^c_{c'}\right) } 
e^{-\alpha (\vec{q}^c)^2}
\nonumber\\
&=-2\lambda \sum_c \int_{1/\Lambda^2}^{\infty} d\alpha e^{-\alpha m^2} \int_{0}^{2\pi} \dfrac{d\beta}{2\pi} e^{i\beta p_{c}} e^{-\alpha p_{c}^{2}}\prod_{c' \neq c} \sum_{q^c_{c'}} 
e^{i\beta q^c_{c'}} e^{-\alpha  (q^c_{c'})^{2}}.
\end{align}
In the last equality, we introduced an integral representation of the Kronecker delta. Now, we can turn the argument of the exponent into a perfect square
and  obtain, for $\vec p \in \cP$
\begin{align}
\Gamma_2^{melo,1}(\vec{p})&=-2\lambda \sum_c\int_{1/\Lambda^2}^{\infty} d\alpha e^{-\alpha m^2} \int_{0}^{2\pi} \dfrac{d\beta}{2\pi} e^{i\beta p_{c}} 
e^{-\alpha p_{c}^{2}}e^{-5\beta^2/4\alpha}\prod_{c' \neq c} \sum_{q^c_{c'}}  e^{-\alpha(q^c_{c'}-i \beta/2\alpha)^{2}}\\
&\sim -2\lambda \sum_c \int_{1/\Lambda^2}^{\infty} d\alpha e^{-\alpha m^2} \int_{0}^{2\pi} \dfrac{d\beta}{2\pi} e^{i\beta p_{c}} e^{-\alpha p_{c}^{2}}
e^{-5\beta^2/4\alpha}\left(\frac{\pi}{\alpha}\right)^{5/2},
\end{align}
in the $\alpha \rightarrow 0$ limit. This identifies the divergent behavior of this expression. These divergencies come from the neighborhood $\alpha=0$, and using the distributional expansion 
\begin{equation}
e^{-\beta^2/4\alpha}=\sqrt{4 \pi \alpha }[ \delta (\beta)+\alpha \delta^{''}(\beta) ]+ \mathcal{O}(\alpha^5/2),
\end{equation}
we obtain:
\begin{align}
\Gamma_2^{melo,1}(\vec{p})&=-2\lambda \sum_c \int_{1/\Lambda^2}^{\infty} d\alpha e^{-\alpha m^2} \int_{0}^{2\pi} \dfrac{d\beta}{2\pi} e^{i\beta p_{c}} e^{-\alpha p_{c}^{2}} \sqrt{4 \pi \alpha }[ \delta (\beta)+\alpha \delta^{''}(\beta) ] \left(\frac{\pi}{\alpha}\right)^{5/2} +\mathcal{O}(1/\Lambda) \nonumber\\
&=-\dfrac{2\lambda\pi^2}{\sqrt{5}}\sum_c \int_{1/\Lambda^2}^{\infty} d\alpha e^{-\alpha m^2}e^{-\alpha p_{c}^{2}}\left( \frac{1}{\alpha^2}-p_c^2\dfrac{1}{5\alpha}\right) +\mathcal{O}(1/\Lambda). \label{equ40}
\end{align}

The asymptotic expansion of this expression at large $\Lambda$ is now easy to find using integrating by parts
\begin{align}
\mathcal{I}&=\int_{1/\Lambda^2}^{\infty} d\alpha e^{-\alpha m^2}e^{-\alpha p_{c}^{2}}\frac{1}{\alpha^2}.
 \nonumber \\
&=\int_{1/\Lambda^2}^{\infty} d\alpha e^{-\alpha m^2}\frac{1}{\alpha^2}-p_c^2\int_{1/\Lambda^2}^{\infty} d\alpha e^{-\alpha m^2}\frac{1}{\alpha}+\mathcal{O}(1/\Lambda)
 \nonumber \\
&=\Lambda^2e^{-m^2/\Lambda^2}-(p_c^2+m^2)\int_{1/\Lambda^2}^{\infty} d\alpha e^{-\alpha m^2}\frac{1}{\alpha}+\mathcal{O}(1/\Lambda).
\end{align}
The divergent part of the last integral is at most logarithmic near zero. Thus:
\begin{equation}
\int_{1/\Lambda^2}^{\infty} d\alpha e^{-\alpha m^2}\frac{1}{\alpha}=A \ln(\Lambda)+\mathcal{O}(1/\Lambda),
\end{equation}
and it suffices to determine A. Differentiating with respect to $ \Lambda $ and identifying the singularity in the two expressions, we find
\begin{align}
\dfrac{d}{d\Lambda}\int_{1/\Lambda^2}^{\infty} d\alpha e^{-\alpha m^2}\frac{1}{\alpha}&=\dfrac{2}{\Lambda^3}e^{-m^2/\Lambda^2}\Lambda^2=\dfrac{2}{\Lambda}e^{-m^2/\Lambda^2} \nonumber \\
&=\dfrac{2}{\Lambda}+\mathcal{O}(1/\Lambda^3)=A(\beta)\dfrac{1}{\Lambda}+\mathcal{O}(1/\Lambda^2)\Rightarrow A=2,
\end{align}
and we obtain the following divergent part:
\begin{equation}
\mathcal{I}=\Lambda^2-2(p_c^2+m^2)\ln(\Lambda)+\mathcal{O}(1).
\end{equation}

Returning to \eqref{equ40}, we find then
\begin{align}
\Gamma_2^{melo,1}(\vec{p})(\vec{p})&=-\dfrac{2\lambda\pi^2}{\sqrt{5}}\sum_c\left( \Lambda^2-2(p_c^2+m^2)\ln(\Lambda)-\dfrac{2}{5}p_c^2\ln(\Lambda)\right)+\mathcal{O}(1/\Lambda) \nonumber \\
&=-\dfrac{12\lambda\pi^2}{\sqrt{5}}\left( \Lambda^2-2m^2\ln(\Lambda)\right)+\dfrac{24\lambda\pi^2}{5\sqrt{5}}\ln(\Lambda)\vec{p}^2
+\mathcal{O}(1/\Lambda). \label{equ70}
\end{align}
and comparing with \eqref{equeff2pt} we conclude that at one loop
\begin{align}
\delta_ZZ&=\dfrac{24\lambda\pi^2}{5\sqrt{5}}\ln(\Lambda), \label{equ71}\\
\delta_{m^2}m^2 &=-\dfrac{12\lambda\pi^2}{\sqrt{5}}\left(\Lambda^2-2m^2\ln(\Lambda)\right).
\end{align}

\subsection{Coupling constant renormalization and asymptotic freedom}

In this section we examine how the coupling changes along the RG trajectory i.e. going towards the IR.. 
Equations \eqref{equ70}-\eqref{equ71} gives us the coefficient $\gamma $ at first order:
\begin{equation}
\gamma= - \Lambda\dfrac{d}{d\Lambda} \ln{Z} =-\dfrac{24\lambda\pi^2}{5\sqrt{5}}.
\end{equation}

It remains now to evaluate the melonic monocolor four-point function at one loop,
$\Gamma^{melo,1}_{4,mono} $.

The contributing diagram is sketched in Figure \ref{figoneloopfour} (shown with its four external arcs). 
\begin{figure}[here]
\begin{center}
\includegraphics[scale=0.8]{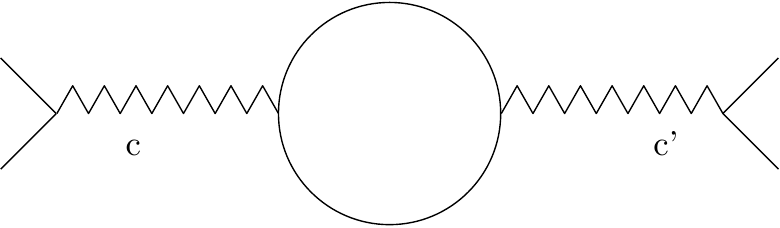} 
\caption{One loop melonic 4-point function}\label{figoneloopfour}
\end{center}
\end{figure}

The total contribution of this diagram is
\begin{equation}
\mathcal{I}':=-\frac{1}{2}2\lambda \sum_c\sum_{q^c_{c'}\, c'\neq c} \dfrac{\delta (\sum_{c'} q^c_{c'})}{[(\vec{q}^c)^2+m^2]^2}=\dfrac{d}{dm^2}\lambda \sum_c\sum_{q^c_{c'}\, c'\neq c} \dfrac{\delta (\sum_{c'} q^c_{c'})}{(\vec{q}^c)^2+m^2},
\end{equation}
and we can deduce the divergent part of $\mathcal{I}'$ (in the same notations as in the previous section):
\begin{equation}
\mathcal{I}'=-\dfrac{12\lambda\pi^2}{\sqrt{5}}\ln(\Lambda) + O(1/\Lambda).
\end{equation}

The last thing to evaluate are the symmetry factors. We have four ways to connect the external fields in an amputated vertex, and two ways to connect these contracted vertices for each of the 6 colors. The expression of the four point function at zero momentum is then ultimately, to the (leading) one-loop order: 
\begin{equation}
\Gamma^{melo,1}_{4,mono} = 6\times 4\left(-\lambda+ \dfrac{2\lambda^2\pi^2}{\sqrt{5}}\ln(\Lambda)\right) + O(1/\Lambda) .
\end{equation}
Returning to equations \eqref{callansym} and \eqref{betafunction} we get:
\begin{equation}
\dfrac{2\lambda^2\pi^2}{\sqrt{5}}-\beta(\lambda)-\dfrac{48\lambda^2\pi^2}{5\sqrt{5}}=0,
\end{equation}
which implies immediately,
\begin{equation}
\Lambda \dfrac{d\lambda}{d\Lambda}=\beta(\lambda)=-\dfrac{38\lambda^2\pi^2}{5 \sqrt{5}}. \label{asymfree}
\end{equation}

The minus sign is fundamental. It means that the bare coupling constant decreases when the ultraviolet cutoff increases. The theory is therefore asymptotically free, thus consistent at the perturbative level, like the familiar non-Abelian gauge theories of the standard model.

We now discuss what happens beyond one loop.

\subsection{Counterterms and renormalons}

Renormalized amplitudes $A_{R}(\mathcal{G})$ can be explicitly written in terms of Zimmermann's forest formula
\begin{equation}
{\cal A}_{R}(\mathcal{G})=\sum_{\mathcal{F}\in D(\mathcal{G})}\prod_{\gamma \in \mathcal{F}}(-\tau^{*}_{\gamma})A(\mathcal{G}), \label{Zifor}
\end{equation}
where $\tau^{*}$ is an operator which performs explicitly the subtraction of the counter-term 
and $D(\mathcal{G})$ is the set of all divergent forests of $\cG$.
However such renormalized amplitudes suffer from the problem of \emph{renormalons}.
Indeed they can grow as  $n!$ with the number $n$ of vertices in $\mathcal{G}$. 	
This problem exists also in our model, and even in its melonic approximation.
Consider indeed the two point subgraph of Figure \ref{figrenormalon}; made of an arbitrarily large monocolor chain
of $n$ simple loop vertices with two arcs, ending on a leaf with a single arc. All wavy lines have same color $c$ and carry the same
momentum $p_c$. 

\begin{figure}[here]
\begin{center}
\includegraphics[scale=1]{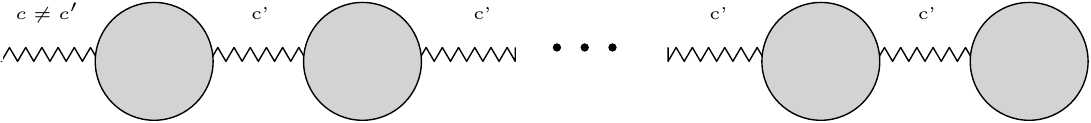} 
\caption{Typical melonic graph with renormalon effect}\label{figrenormalon}
\end{center}\end{figure}

Because the renormalized four point function, hence the 
renormalized loop vertex with two arcs, behaves as $\log(p_c)$ at large $p_c$, inserting such a chain on a convergent loop
in a convergent melonic vertex function will lead to a very large sum over $p_c$ 
which typically can behave at large $n$ as
\begin{equation}
\sum_{p_c \in {\mathbb Z}} [\log{p_c}]^n  \dfrac{1}{p_c^2 +m^2} \sim K^n n! \;\; 
\end{equation}
for some constant $K$. This is the renormalon problem. 

Such renormalons in fact come entirely from the counterterms in Zimmermann's formula \cite{Rivasseau:1991ub}. 
More precisely in \eqref{Zifor} the counter-terms are subtracted, so to speak, blindly with respect to internal scale integrations. 
But a divergent subgraph looks like its counter term only when its internal scales are higher than the  scale its external lines, hence when it
is a $\cG^k_i$ in some attribution $\mu$ in the language of the previous section (locality principle).
Counter-terms in \eqref{Zifor} not only subtract these dangerous contributions, but also the inoffensive parts 
in which the internal lines of the divergent subgraph 
have lower scale than the external lines. It is exactly these unnecessary subtractions which give rise to the renormalons. 
Hence, although the standard renormalization procedure 
eliminates all ultraviolet divergencies from any Feynman amplitude, such renormalized amplitudes 
are so big that we cannot use them directly
to sum even the melonic approximation to our theory.

\subsection{The effective amplitudes}

The effective series is a more physical way to compute perturbation theory, and a natural solution to the renormalon problem 
when the theory is asymptotically free \cite{Rivasseau:1991ub}. The basic idea is to renormalize in the Wilsonian spirit, namely step by step, expanding 
in a whole sequence of effective couplings rather than in the single renormalized coupling. 
Consider a graph $\mathcal{G}$ and its bare amplitude $A_{\mu}(\mathcal{G})$ with scale attribution $\mu$ as defined in the previous section.  
There are some $\cG^k_i$  subgraphs which are divergent ($\omega (\cG^k_i)\ge 0$). They form a forest 
$D_{\mu}(\mathcal{G})$ (because it is a subset of the Gallavotti-Nicol\`o  tree containing all $\cG^k_i$ high subgraphs). 
The effective amplitude ${\cal A}^{eff}(\mathcal{G})$ is defined by
\begin{equation}
{\cal A}^{eff}(\mathcal{G})  = \sum_\mu {\cal A}_\mu^{eff}(\mathcal{G}) , \;\; {\cal A}^{eff}_{\mu}(\mathcal{G}):=\prod_{\gamma \in D_{\mu}(\mathcal{G})}(1-\tau^{*}_{\gamma})A_{\mu}(\mathcal{G}). \label{equeff1}
\end{equation}
Comparing with \eqref{Zifor}, we see such amplitudes are very different from the renormalized ones.
Because in \eqref{equeff1} all divergent high graphs are subtracted, effective amplitudes, like renormalized ones,
have a finite limit when the ultraviolet cutoff is removed. However unlike renormalized amplitudes, effective amplitudes are free of renormalons 
\cite{Rivasseau:1991ub,Carrozza22013}. More precisely

\begin{theorem}  
The effective amplitude $\cA_{eff}(\mathcal{G})$ for a graph $\mathcal{G}$ with $V(\mathcal{G})$ internal wavy lines is 
uniformly exponentially bounded in term of its size, hence for some constant $K$ 
\begin{equation}
|A_{eff}(\mathcal{G})|\leq K^{V(\mathcal{G})}.
\end{equation}
\label{theonoreno}
\end{theorem}
\noindent{\bf Proof} {\it (sketched)}
Renormalization operators exactly act on the divergent subgraphs $\cG^k_i$ only. Taylor expanding and using the 
condition that external legs of $\cG^k_i$ have all lower scales than any internal line, they transform their
divergent degree into an effective degree $\omega' \le -1$. The rest of the argument to bound the 
sum over $\mu$ then follows exactly the proof of Theorem \ref{theowein}. \qed

Hence effective amplitudes are better building blocks than either bare or renormalized amplitudes to understand the ultraviolet
limit of the theory. It remains to relate them to the initial theory.
Consider the bare power series defined by:
\begin{equation}
G^{\Lambda}_{2N}=\sum_{\mathcal{G}, \mu} \frac{1}{s(\mathcal{G})} (-\lambda )^{V( \cG)} \cA_{\mu}(\mathcal{G}), \label{equbare}
\end{equation} 
where attributions $\mu$ are summed with cutoff $\Lambda = M^{-2\rho}$ (hence by \eqref{cutoffsha} every scale satisfies $1 \le i_\ell \le \rho$)
the amplitudes $\cA_{\mu}$ are computed with bare propagators and $\lambda$ is the bare coupling. It has obviously
no ultraviolet limit. But we have the following key theorem \cite{Rivasseau:1991ub, GalNic, Carrozza22013}:
\begin{theorem}
\textbf{(Effective expansion)} The series \eqref{equbare} can be reshuffled as 
a multi-series with effective couplings and effective amplitudes:
\begin{equation}
G^{\Lambda}_{2N}=\sum_{\mathcal{G}, \mu} \frac{1}{s(\mathcal{G})}\left(\prod_{v \in \mathcal{V}(\mathcal{G})} (-\lambda^{(\Lambda)}_{i_v(\mu)}) \right) 
\cA_{\mu}^{eff}(\mathcal{G}),
\end{equation}
where $i_v(\mu) =\sup \{i_\ell, \ell \; {\rm hooked\, to}\, v\}$, and
the effective couplings $\lambda^{(\Lambda)}_{i_v(\mu)}$ and the effective propagators $C^{eff}_i$
occurring for lines of scale $i$ in the amplitude  $\cA_{\mu}^{eff}(\mathcal{G})$
obey the inductive relations \eqref{recu1}-\eqref{recu2} below.
Moreover, defining the renormalized coupling by $\lambda_r := \lambda_{0}$ and the renormalized propagators $C_r$
by inverting \eqref{equeff2pt}, and reshuffling the effective series in terms of the single renormalized 
coupling $\lambda_r$ with renormalized propagators $C_r$, we recover exactly the renormalized series. 
\end{theorem}

In particular $\lambda^{(\Lambda)}_{\rho}$ is the bare coupling, and $\lambda^{(\Lambda)}_{0}$ is the renormalized one. The other
couplings $\lambda_i$ for $0< i < \rho$ describe the RG trajectory in between these extremal values. \\

\noindent{\bf Proof} {\it (sketched)}
We recall only the main steps in the proof; more details can be found in \cite{Rivasseau:1991ub,Carrozza22013}. 

The proof is inductive, working from the high scales towards the lower ones. The initial step $i=\rho$ starts with the bare series. 
At step number $i$ we suppose we have defined the effective expansion with 

\begin{itemize}
\item effective couplings $\lambda_{j}$  for vertices with highest scale
$j>i$ and $\lambda_{i+1}$ for all vertices with highest scale $j\le i$; 

\item effective propagators $C_j$ for lines with indices $j >i$ and $C_{i+1}$ for all lines with indices $j\le i$, 

\item 
effective amplitudes ${\cal A}^{eff,i+1}(\mathcal{G}) $ 
with subtractions $\prod_{\gamma \in D^{i+1}_{\mu}(\mathcal{G})}(1-\tau^{*}_{\gamma})$, where 
$D^{i+1}_{\mu}$ is the forest of all divergent $\cG_{j}^k$ with $j > i$.
\end{itemize}

We define the next coupling $\lambda_{i}$ and propagator $C_i$ by considering in $\mu$ the scale number $i$.
Adding and subtracting the counter-terms in $D^{i}_{\mu}\setminus D^{i+1}_{\mu}=\{\cH \in D_\mu(\mathcal{G}) | 
\inf i_\ell = i\} , \ell \in \cH$,
we write
\begin{equation}
\cA^{eff,i+1}_{\mu}(\mathcal{G}):=\prod_{\cH\in D^{i}_{\mu}\setminus D^{i+1}_{\mu}}
 [(1-\tau^{*}_{\cH})+\tau^{*}_{\cH}]\prod_{\gamma \in D_{\mu}^{i+1}}(1-\tau^{*}_{\gamma})\cA_{\mu}^{eff,i+1}(\mathcal{G}),
\end{equation}
and we expand the product over $\cH\in D^{i}_{\mu}\setminus D^{i+1}_{\mu}$.
The operators $(1-\tau^{*}_{\cH})$ will generate the next layer of subtraction in the formula to
change the subtraction operations of $\cA^{eff,i+1}_{\mu}(\mathcal{G})$ into those of $A^{eff,i}_{\mu}(\mathcal{G})$.
The counterterms $+\tau^{*}_{\cH}$ are then associated to collapsed graphs $\cG/\cH$ 
in which $\cH$ has been collapsed to a vertex (if $N(\cH) =2$) or to a mass
or a wave function insertion (if $N(\cH)=1$). Collecting these pieces and rearranging them according to the collapsed graph rather
than to the initial graph defines an (infinite series) redefinition of the couplings hooked to 
vertices with highest line of slice $j< i$   and of the propagators with scale $j<i$, which become respectively $\lambda_i$
and $C_i$. Hence the new effective coupling is
\begin{equation}
-\lambda_i  = -\lambda_{i+1} + \sum_{\cH \vert \; N(\cH =2) , \inf_{\ell \in \cH} i_\ell = i  }  \tau^{*}_{\cH}  \cA_{\mu}^{eff,i+1}(\mathcal{H})
\label{recu1}
\end{equation}
and the new propagator is 
\begin{equation}
C_i  = C_{i+1} + \sum_{\cH \vert \; N(\cH =2) , \inf_{\ell \in \cH} i_\ell = i  }  \tau^{*}_{\cH}  \cA_{\mu}^{eff,i+1}(\mathcal{H}).
\label{recu2}
\end{equation}
Remark we can omit in these definitions that $\cH$ is divergent, since $ \tau^{*}_{\cH} = 0$ if $\cH$ is convergent. Remark
also that $\cH$ in \eqref{recu2} is connected but can be one particle reducible
and that to update the effective mass and effective Laplacian normalization in $C_r$ from $i+1$ to $i$
requires to analyze  \eqref{recu2} in terms of
the one-particle irreducible self-energy (see \eqref{equeff2pt}).
Finally remark also that such recursive equations are \emph{non-Markovian}. By this we mean that the 
effective coupling $\lambda_{i}$ is itself a multi-series in the sequence
of all effective couplings $\lambda_\rho, \cdots , \lambda_{i+1}$, 
Any attempt to rewrite it in terms of the single coupling $\lambda_{i+1}$ would automatically reintroduce 
the renormalon problem.
\qed

Thanks to Theorem \ref{theonoreno} the effective expansion is therefore able to define the theory provided all couplings on the trajectory 
from $\lambda_\rho$ to $\lambda_0 = \lambda_r$ are uniformly bounded by a sufficently small constant, 
and the number of graphs is not too big. This is the case when

\begin{itemize}

\item the theory is asymptotically free or asymptotically safe in the ultraviolet regime,

\item the set of graphs considered does not proliferate more than exponentially with size $n$.

\end{itemize}
Planar ``wrong sign" $\phi^4$   \cite{'tHooft:1982cx} or the Grosse Wulkenhaar model \cite{Grosse:2004yu} satisfy these two conditions. Since melonic graphs,
like trees, obviously proliferate no more than exponentially 
in size and since our theory is asymptotically free, its melonic approximation also satisfy both conditions. Hence
the effective expansion allows to define non perturbatively this
melonic approximation, in fact for any Green function $G^{melo}_{2N}$ or vertex function $\Gamma^{melo}_{2N}$.

\section{Melonic Equations 
}\label{sec:closed}
In this section, we establish a closed equation for the melonic two-point vertex function, and an 
equation  expressing the melonic four-point vertex function in terms of the two-point one. Combining this with the effective bounds of the previous section
we shall prove existence and unicity of the solution of these equations
at small renormalized coupling.

\subsection{Bare Equations}

Let us start with the two-point vertex function or self-energy. The relationship between the Green function $G_2$ 
and the self-energy can be graphically represented as:
\begin{figure}[h]\begin{center}
\includegraphics[scale=.9]{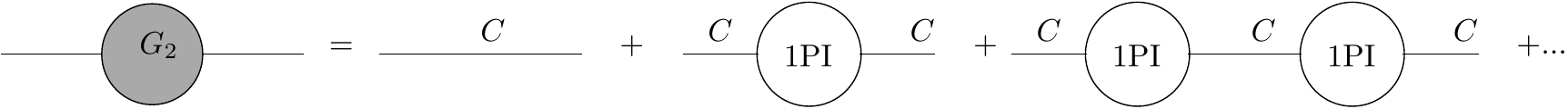}
\caption{Decomposition of the 2-point function}
\end{center}\end{figure}

\noindent
corresponding to the functional relationship:
\begin{equation}
G_2(\vec{p})=C+C\Gamma_2C+C\Gamma_2C\Gamma_2C+\cdot\cdot\cdot=\dfrac{C}{1-\Gamma_2 C}
=\dfrac{\delta(\sum_c p_c)}{\vec{p}^2+m^2-\Gamma_2(\vec{p})}. \label{onepistruct}
\end{equation}

We want to restrict now this relationship to the melonic approximation.
\eqref{goodeq1}
expressed $\Gamma^{melo}_2$ as a sum of trees
in the intermediate field representation. Focusing on the root of the tree,  we can amputate the unique ciliated vertex
into two trivial half-lines (this wont be possible if there were wavy loops). Detailing the loop vertex at the other end of the unique wavy
line of the tree connected to the ciliated vertex leads to the following graphical representation
of  $\Gamma^{melo}_2$
\begin{figure}[here]
\begin{center}
\includegraphics[scale=1]{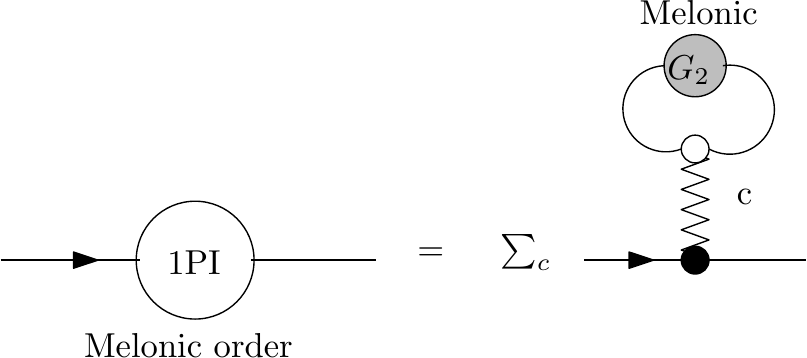} 
\caption{1PI melonic two point function}
\end{center}\end{figure}
\noindent

where we sum over all possible colors for the root wavy line. Combining with \eqref{onepistruct} we get
\begin{equation}
\Gamma_2^{melo}(\vec{p})=-2\lambda \sum_c \sum_{{q}^c_{c' \, c'\neq c}} G_{2}(\vec{q}^c)=
-2\lambda \sum_c \sum_{{q}^c_{c' \, c'\neq c}}
\dfrac{\delta(\sum_c {q^c_{c'})}}{(\vec{q}^c)^2+m^2-\Gamma_2^{melo}(\vec{q}^c)},
\end{equation}
where the vector $\vec{q}^c$ was defined in the previous section. This is a closed equation for the melonic  self energy. 
Using Theorem \ref{goodth}, it writes in terms  of the function $f$ as
\begin{equation}
f(p)= 2 \sum_{q^1_{c'}, \, 2\leq c' \leq 6}
\dfrac{\delta(\sum_{c' } q^1_{c'}   )   }
{( \vec{q}^1)^2 +m^2 + \lambda \sum_{c'} f ( q^1_{c'}  )   }. \label{bareclosed}
\end{equation}
where we recall that $q^1_{c'}= \{  p, q^1_2  , \cdots q^1_6\}$ is a function of $p$.

Turning now to the melonic four-point vertex function, 
we can draw the two end vertices as in Figure \ref{twovertfig}.
\begin{figure}[h]
\begin{center}
\includegraphics[scale=.8]{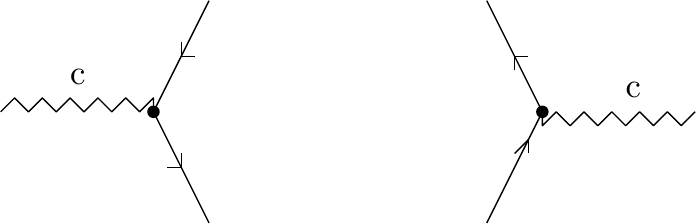} \caption{End vertices of $\Gamma_{4,int}^{melo}$}\label{twovertfig}
\end{center}
\end{figure}

Using the results of section
\ref{sectionreg}
on the monocolored tree structure of $\Gamma_{4}^{melo}$ and taking care of the combinatorics we can write
\begin{equation}
\Gamma_{4}^{melo}(\vec{p}_1,...\vec{p}_4)= -4 \lambda \sum_c [   1 - \lambda g_{int}( p_c  )]Sym \mathcal{M}^{(c)}_{\vec{p}_1,...\vec{p}_4},
\end{equation}
where $\mathcal{M}^{c}$ is define by $\Tr_{b_c}(\phi,\bar{\phi})=:\sum_{\{\vec{p}_i\}}\mathcal{M}^{(c)}_{\vec{p}_1,\vec{p}_2,\vec{p}_3\vec{p}_4}\phi_{\vec{p}_1}\bar{\phi}_{\vec{p}_2}\phi_{\vec{p}_3}\bar{\phi}_{\vec{p}_4}$, 
\begin{equation*}
Sym \mathcal{M}^{(c)}_{\vec{p}_1,...\vec{p}_4}:=\frac{1}{2}\big(\mathcal{M}^{(c)}_{\vec{p}_1,\vec{p}_2,\vec{p}_3\vec{p}_4}+\mathcal{M}^{(c)}_{\vec{p}_3,\vec{p}_2,\vec{p}_1\vec{p}_4}\big),
\end{equation*}
and $\Gamma_{4,int}^{melo}(\vec{p}) := \sum_c  g_{int}( p_c  )Sym \mathcal{M}^{(c)} $ is the simple loop integral with two arcs corresponding to Figure \ref{1PImelonicfoufig}.
\begin{figure}[h]
\begin{center}
\includegraphics[scale=.6]{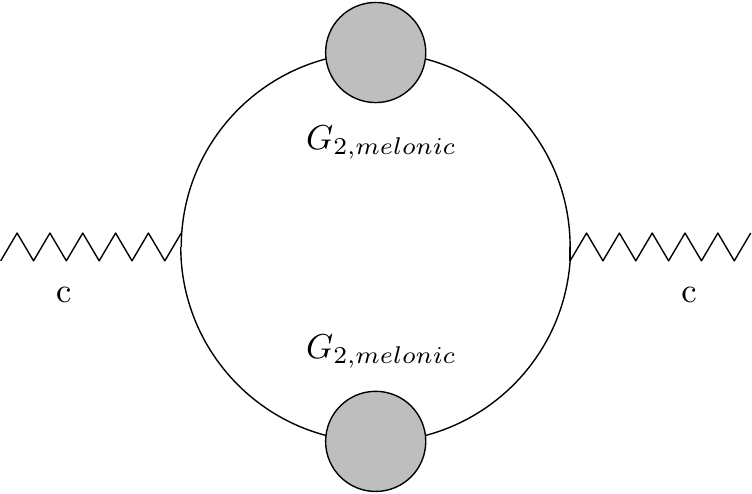} \caption{A melonic two point insertion}\label{1PImelonicfoufig}
\end{center}
\end{figure}

Hence
\begin{equation}
g_{int}(p)= \sum_{q^c_{c'\neq c}} \dfrac{\delta(\sum_{c'} q^c_{c'}) }{\Big[ (\vec{q}^c)^2+m^2+ \lambda \sum_{c'} f(q^c_{c'})\Big]^2}. 
\label{gint}
\end{equation}
Using Theorem \ref{goodth} this means
the following relation between $f$ and $g$
\begin{equation}
g(p)= 4 \bigg( 1 -\sum_{q^c_{c'\neq c}} \dfrac{\delta(\sum_{c'} q^c_{c'}) }{\Big[ (\vec{q}^c)^2+m^2 + \lambda  \sum_{c'} f(q^c_{c'})\Big]^2}  \bigg) . 
\label{barefour}
\end{equation}

\subsection{Renormalized Equations}
In this subsection we give the renormalized version of the previous equations. 
We rewrite  \eqref{onepistruct} according to section \ref{sectionren} as
\begin{equation}
G_2(\vec{p})=\dfrac{\delta(\sum_c p_c)}{Z\vec{p}^2+Z_{m^2}m^2_{r}-\Gamma_2(\vec{p})}=\dfrac{\delta(\sum_c p_c)}{\vec{p}^2+m^2_{r}+\delta_{Z}\vec{p}^2+\delta_{m^2}m^2_{r}-\Gamma_2(\vec{p})}
\end{equation}
Next, we impose the following useful renormalization conditions :
\begin{align}
\delta Z&:=\dfrac{d\Gamma_2}{d\vec{p}^2}\vert_{\vec{p}=\vec{0}}\\
\delta_{m^2}&:=\Gamma_2(\vec{p}^2=0).
\end{align}
Note about this expression that, strictly speaking, the function $\Gamma_2$ is a function on $\mathbb{Z}^d$, and the derivation operation makes no sense. In the last expression, the derivative can be viewed as a new function on $\mathbb{Z}^d$, obtained from the first by analytic prolongation on $\mathbb{R}^d$ (with the preamble remark that the function on $\mathbb{Z}^d$ admit a natural prolongation on the continuous space), computation of the derivative of this new function, and finally restriction to the subset $\mathbb{Z}^d \subset \mathbb{R}^d$.\\

The renormalized function is therefore obtained from the previous equation and the renormalization conditions, by subtracting its value at $\vec{p}^2=0$ and its first derivative at the same point:
\begin{equation}
\Gamma_{2,r}(\vec{p}):=\Gamma(\vec{p}^2)-\Gamma_2(\vec{0})-\vec{p}\,^2\dfrac{d\Gamma_2}{d\vec{p}\,^2}\vert_{\vec{p}=\vec{0}},
\end{equation}
A similar equation relates $\Gamma_{2,r}^{melo}(\vec{p})$ to $\Gamma_{2,r}^{melo}(\vec{p})$,
and by the same argument that in the previous section, we obtain the following closed equation for the renormalized self-energy:
\begin{align}
&\Gamma_{2,r}^{melo}(\vec{p})= -2
\lambda_{r}  \sum_c \sum_{q^c_{c'\neq c}}\bigg[ \dfrac{\delta(\sum_{c'} q^c_{c'}) }{(\vec{q^c})^2+m^2_{r}-\Gamma_{2,r}^{melo}(\vec{q^c})}\\\nonumber
&-\dfrac{\delta(\sum_{c'} q^c_{c'}) }{(\vec{q^c})^2+m^2_{r}-\Gamma_{2,r}^{melo}(\vec{qc})}\vert_{p_c=0}-\sum_c p_{c}^{2}\dfrac{d}{dp_{c}^{2}}
\dfrac{\delta(\sum_{c'} q^c_{c'}) }{(\vec{q^c})^2+m^2_{r}-\Gamma_{2,r}^{melo}(\vec{q^c})}\vert_{p_c=0}\bigg].
\end{align}
We obtain an equation for $f_r$, the renormalized function $f$
such that 
\beq \Gamma_{2,r}^{melo} (\vec p) = - \lambda_r  \sum_c   f_r (p_c),
\eeq  
namely
\bea
&f_r(p_c)=
2\sum_{q^1_{c'}, \, 2\leq c' \leq 6} \bigg[ \dfrac{\delta(\sum_{c' } q^1_{c'}   ) }{( \vec{q}^1)^2 +m^2_{r}+ \lambda_r \sum_{c'} f_r( q^1_{c'}  ) }
\\\nonumber
&-\dfrac{\delta(\sum_{c' } q^1_{c'}   ) }{( \vec{q}^1)^2 +m^2_{r} + \lambda_r \sum_{c'} f_r( q^1_{c'}  ) }\vert_{p_c=0} 
- p_{c}^{2}\dfrac{d}{dp_{c}^{2}}\dfrac{\delta(\sum_{c' } q^1_{c'}   ) }{( \vec{q}^1)^2 +m^2_{r} + \lambda_r \sum_{c'} f_r( q^1_{c'}  )  }\vert_{p_c=0}\bigg].
\label{baretworen}
\eea
The renormalized equation corresponding to
\eqref{barefour}
follows in the same way. Setting
\beq
\Gamma_{4,mono,r}^{melo} (\vec p, \vec p\,') = - \lambda_r  \sum_c  \delta(p_c, p'_c) g_r (p_c)
\eeq
(compare with \eqref{goodeq}) we have the renormalized version of \eqref{gint}
\begin{equation}
g_{int,r}(p)= \sum_{q^c_{c'\neq c}}\bigg[  \dfrac{\delta(\sum_{c'} q^c_{c'}) }{\Big[ (\vec{q}^c)^2+m_r^2+ \lambda_r \sum_{c'} f_r(q^c_{c'})\Big]^2}
- \dfrac{\delta(\sum_{c'} q^c_{c'}) }{\Big[ (\vec{q}^c)^2+m_r^2+ \lambda_r \sum_{c'} f_r(q^c_{c'})\Big]^2}\vert_{p=0}\bigg],
\label{gintren}
\end{equation}
and the renormalized version of \eqref{barefour} 
\begin{equation}
g_r(p)= 4 \bigg( 1 -\sum_{q^c_{c'\neq c}}
\bigg[  \dfrac{\delta(\sum_{c'} q^c_{c'}) }{\Big[ (\vec{q}^c)^2+m_r^2 + \lambda_r \sum_{c'} f_r(q^c_{c'})\Big]^2}  -
\dfrac{\delta(\sum_{c'} q^c_{c'}) }{\Big[ (\vec{q}^c)^2+m_r^2 + \lambda_r \sum_{c'} f_r(q^c_{c'})\Big]^2}\vert_{p=0}\bigg]
\bigg) . 
\label{barefourren}
\end{equation}

\subsection{Existence and Unicity}
The previous closed equations define, in principle, the renormalized melonic vertex functions. Neither the existence nor the unicity of their solutions, however, are obvious at all, since the bare equations do not have ultraviolet limit and the renormalized ones typically have zero
convergence radius in $\lambda_r$ because of renormalons (except at very special values such as zero external momenta). 
But we can expand these equations according the multiscale expansion of Section
\ref{sectionreg}
and reshuffle them in terms of the 
effective amplitudes and effective constants $\lambda_i$ of Section
\ref{sectionren}. Subtractions in loop sums such as those of \eqref{baretworen} and \eqref{barefourren} will then occur only when the external momentum $p_c$ has scale strictly lower
than the one of $\vec q\,^c$
and the coupling $\lambda_r $ will be replaced by the effective coupling corresponding to the scale of $\vec q\,^c$.

Expanding in a multiseries for all couplings gives therefore an effective expansion with

\begin{itemize}

\item at most $(K_1)^n$ graphs at order $n$, since as well known, trees proliferate only exponentially
in their number of vertices,

\item effective melonic amplitudes bounded by $(K_2)^n$ by Theorem \ref{theonoreno} (which applies to \emph{any} effective amplitude, hence in particular to the melonic ones),

\item effective constants all bounded by the last one $\lambda_r$ because of asymptotic freedom \eqref{asymfree}.

\end{itemize}

Hence this effective melonic expansion converges and defines a unique solution of the renormalized equations for $0\le \lambda_r < (K_1K_2)^{-1}$.
As usually for  flow equations such as  \eqref{asymfree}, the solution is in fact analytic in $\lambda_r$ in a disk tangent to the real axis,
with uniform Taylor remainder estimates at order $s$ in $K^s s!$ \cite{Rivasseau:1991ub}. 
We leave the details to the reader, but have no doubt that the unique solution sum of the effective series is therefore the \emph{Borel sum}
of the renormalized expansion for the melonic vertex functions $\Gamma^{melo}_{2N,r}$, and that this holds 
not just for $N=1$ and 2 but for \emph{any} number $2N$ of external arguments. This completes the
control of the melonic sector of the theory:

\begin{theorem}
The effective expansions of the renormalized melonic vertex functions converge for $0 \leq \lambda_r < K^{-1}$
to the  Borel sum of their renormalized expansions.
\end{theorem}

It is tempting to believe that like for tensor models \cite{critical},
for $\vert \lambda_r\vert $ large enough we reach singularities at which phase transitions occur, but this is left to future analytic and numerical study.

\section{Conclusion}
We have studied a simple Abelian TGFT of rank 6 with quartic melonic interactions.
We defined its intermediate field representation and used it, 
together with a multi-scale analysis, to prove its renormalizability, to compute its beta function and to check its asymptotic freedom. 
We have defined the effective expansion of the model and established uniform exponential
upper bounds on effective melonic amplitudes. Finally we wrote a closed equation for the melonic approximation to the two-point and four-point 
vertex functions
and using the effective expansion we proved that it admits a unique solution for small enough stable renormalized coupling. 

Next steps in the analysis of the model might be the numerical analysis of the RG flow along the lines of \cite{Benedetti:2014qsa} and a full constructive analysis
(including the non-melonic sector) of this model. The latter would require a non-trivial extension of the techniques of \cite{Delepouve:2014hfa}, but may be tractable thanks to the vector-like nature
of the intermediate field. 


\end{document}